\documentclass[11pt]{article}

\newlength{\minipagewidth} 
\usepackage{amsmath,amsthm,mathrsfs} 
\usepackage{amssymb} 
\usepackage{a4wide} 
\usepackage{graphicx} 
\usepackage{color} 
\usepackage{epic}
\usepackage{enumerate} 
\usepackage{natbib} 

\newtheorem{theorem}{Theorem}

\newtheorem{algo}[theorem]{Algorithm}

 \newcommand{\bookbox}[1]{
  \par\medskip\noindent
  \framebox[\textwidth]{
    \begin{minipage}{\minipagewidth} {#1}
    \end{minipage} } \par\medskip }

\newcommand\dps {\displaystyle }

\newcommand{\R}{{\mathbb{R}}} 

\newcommand{\E}{{\mathbb{E}}}

\newcommand{\norm}{\mathrm{N}}
\newcommand{\gam}{\mathrm{Gamma}}
\newcommand{\poi}{\mathrm{Poisson}}
\newcommand{\dirich}{\mathrm{Dirichlet}}
\newcommand{\ind}[1]{\mathbf{1}_{\{ #1 \}}}

\begin{document}

  
\title{Free Energy Methods for Bayesian Inference: Efficient Exploration of Univariate Gaussian Mixture Posteriors}
\author{Nicolas Chopin$^{1}$\footnote{Corresponding author: nicolas.chopin@ensae.fr}
, Tony Leli\`evre$^2$ and Gabriel Stoltz$^{2}$\\
  \footnotesize{1: ENSAE-CREST, 3, Avenue Pierre
    Larousse, 92245 Malakoff, France.} \\
  \footnotesize{2: Universit\'e Paris Est, CERMICS, Projet MICMAC Ecole des Ponts ParisTech -
    INRIA} \\
  \footnotesize{6 \& 8 Av. Pascal, 77455 Marne-la-Vall\'ee Cedex 2, France.} \\
}
\date{} 
\maketitle
 
\begin{abstract}
  Because of their multimodality, mixture posterior distributions are
  difficult to sample with standard Markov chain Monte Carlo (MCMC)
  methods. We propose a strategy to enhance the sampling of MCMC in
  this context, using a biasing procedure which originates from
  computational Statistical Physics. The principle is first to choose
  a ``reaction coordinate'', that is, a ``direction'' in which the target
  distribution is multimodal. In a second step, the marginal log-density of
  the reaction coordinate with respect to the posterior distribution is estimated; minus this quantity is called ``free
  energy'' in the computational Statistical Physics literature. To
  this end, we use adaptive biasing Markov chain algorithms which
  adapt their targeted invariant distribution on the fly, in order to overcome
  sampling barriers along the chosen reaction coordinate. Finally, we
  perform an importance sampling step in order to remove the bias and
  recover the true posterior.  The efficiency factor of the importance
  sampling step can easily be estimated \emph{a priori} once the bias
  is known, and appears to be rather large for the test cases we considered.

  A crucial point is the choice of the reaction coordinate.  One
  standard choice (used for example in the classical Wang-Landau
  algorithm) is minus the log-posterior density. We discuss other
  choices. We show in particular that the hyper-parameter that
  determines the order of magnitude of the variance of each component
  is both a convenient and an efficient reaction coordinate.

  We also show how to adapt the method to compute the evidence
 (marginal likelihood) of a mixture model. We
  illustrate our approach by analyzing two real data sets.
  
  \medskip
 
\end{abstract}


\section{Introduction}

Mixture modeling is presumably the most popular and the most flexible
way to model heterogeneous data; see \emph{e.g.}
\cite{TittBookMixtures}, \cite{MclachlanPeelMixtures} and
\cite{Fru:book} for an overview of applications of mixture models.
Due to the emergence of Markov chain Monte Carlo (MCMC), interest in
the Bayesian analysis of such models has sharply increased in recent
years, starting with \cite{DiebRob}. However, MCMC analysis of
mixtures poses several problems \citep{CelHurRob,jasra2005markov}.
In this paper, we focus on the difficulties arising from the
multimodality of the posterior distribution.

For the sake of exposition, we concentrate our discussion on
univariate Gaussian mixtures, but we explain in the conclusion
(Section~\ref{sec:conclusion}) how our ideas may be extended to other
mixture models. The data vector $y=(y_1,\ldots,y_n)$ contains
independent and identically distributed (i.i.d.) real random variables
with density
\begin{equation}
  \label{eq:mixdens}
  p(y|\theta) = \prod_{i=1}^n p(y_i|\theta), \qquad 
  p(y_i|\theta) = \sum_{k=1}^K q_k \, \varphi(y_i;\mu_k,\lambda_k^{-1}),
\end{equation}
where the vector $\theta$ contains all the unknown parameters, 
\emph{i.e.}  the mixture weights $q_1,...,q_{K-1}$, the means
$\mu_1,\ldots,\mu_K,$ the precisions $\lambda_1,\ldots,\lambda_K$, and
possibly hyper-parameters, and $ \varphi(\cdot;\mu,\lambda^{-1})$
denotes the Gaussian density with mean $\mu$ and variance
$\lambda^{-1}$.

This model is not identifiable, because both the likelihood and the
posterior density are invariant by permutation of components, provided
the prior is symmetric. This is the root of the aforementioned
problems. By construction, any local mode of the posterior density
admits $K!-1$ symmetric replicates, while a typical MCMC sampler
recovers only one of these $K!$ copies.  A possible remedy is to
introduce steps that permute randomly the components
\citep{Fru:mixtures}. However, mixture posterior distributions are often
``genuinely multimodal'', following the terminology of
\cite{jasra2005markov}: The number of sets of $K!$ equivalent modes is
often larger than one; see also \citet[][Chap. 6]{BayesianCore} for an
example of a multimodal posterior distribution generated by an identifiable
mixture model, and Section~\ref{sec:meta} of this paper for an example
based on real data. Therefore, and following \cite{CelHurRob} and
\cite{jasra2005markov}, we consider that a minimum requirement of
convergence for a MCMC  sampler is to visit all possible
labelling of the parameter, without resorting to random
permutations.

Inspired by techniques used in molecular dynamics~\citep{LRS10}, 
our aim is to develop samplers that meet this
requirement, using an importance sampling
strategy where the importance sampling function is the marginal distribution of a
well chosen variable. 
The principle of the method is (i) to choose a ``reaction coordinate'',
namely a low-dimensional function of the parameters $\theta$; (ii)
to compute the marginal log-density of this reaction coordinate
(minus this log-density is 
called the ``free energy'' in the molecular dynamics literature); and
(iii) to use the free energy to build a bias for the target of the MCMC sampler,
in order to move more freely between the different modal regions of 
the initial target distribution. More precisely, the biased
log-density is obtained by adding the free energy to the target
log-density; this changes the marginal distribution of the reaction
coordinate into a uniform distribution (within chosen bounds).
At the final stage of the algorithm, the bias can be removed by
performing a simple importance sampling step from the biased posterior distribution to
the original unbiased posterior distribution. Expectations with
respect to the posterior distribution may be computed from the
weighted sample. 
We also derive a method for computing the evidence (marginal likelihood). 

If the reaction coordinate is well chosen, it is likely that a
MCMC sampler targeted at the biased posterior converges to equilibrium much faster
than a standard MCMC sampler, see~\cite{LRS08}.  Two questions are then
in order: How to choose the reaction coordinate? And how to compute
the free energy?

Concerning the choice of the reaction coordinate, the application of
free energy based methods to mixture models is not straightforward. 
In many cases, samplers targeting a mixture posterior distribution are \emph{metastable}:
This term means that  the trajectory generated by the algorithm
remains stuck in the vicinity of some local attraction point for very long times, before moving
to a different region of the accessible space where it again remains stuck.
We show that the degree of metastability of a sampler targeting a mixture posterior
distribution is strongly determined by certain hyper-parameters,
typically those that calibrate in the prior the spread of each
Gaussian component. We show that such hyper-parameters are good
reaction coordinates, in the sense that (i) it is possible to
efficiently compute the associated free energy (by adaptive
algorithms, see below), (ii) the corresponding free energy biased
dynamics explores quickly the (biased) posterior distribution, and
(iii) the points sampled from the biased distribution have
non-negligible importance weights with respect to the original target
posterior distribution. Other reaction coordinates are discussed, such
as the posterior log-density, which is also a good reaction coordinate
(with the problem however that an appropriate range of variation for
this reaction coordinate, which is needed for the method, is difficult
to determine \emph{a priori}). This reaction coordinate is the
standard choice for the Wang-Landau algorithm, see for
instance~\cite{atchade-liu-04}, \cite{Liang05}
and~\cite{Liang2010trajectory} for related works in Statistics.

To compute the free energy, we resort to adaptive biasing
algorithms~\citep{DP01,HC04,marsili-barducci-chelli-procacci-schettino-06,LRS07}. 
In contrast with the adaptive MCMC algorithms usually
considered in the statistical literature (see the review of
\citet{andrieu2008tutorial} and references therein), these adaptive
algorithms modify sequentially the targeted invariant distribution of the
Markov chain, instead of the parameters of the Markov kernel. Such
algorithms were initially designed to compute the free energy in
molecular dynamics; see also~\cite{LRS10} for a recent review of
alternative standard techniques in molecular dynamics for computing
the free energy, such as \emph{e.g.} thermodynamic integration.

It is of course possible to combine our approach with other strategies
aimed at overcoming multimodality, such as tempering methods;
see \emph{e.g.}~\cite{Iba01,Neal:AIS} and~\cite{CelHurRob}.

The paper is organized as follows. Section \ref{sec:model_metastab}
presents the univariate Gaussian mixture model, and the difficulties encountered
with classical MCMC
algorithms. Section~\ref{sec:free_energy_biased_sampling} describes
the method we propose, which is based on free energy biased dynamics.  
Section~\ref{sec:abf-bayes-mixt} explains how to apply
this method to Bayesian inference on Gaussian mixture
models. Section~\ref{sec:examples} illustrates our approach with two
real data-sets.  Section \ref{sec:conclusion} discusses further
research directions, in particular how our approach may be extended to
other Bayesian models.


\section{Scientific context}
\label{sec:model_metastab}

\subsection{Gaussian mixture posterior distribution}
\label{sec:post-distr}
As explained in the introduction, we focus on the univariate Gaussian
mixture model \eqref{eq:mixdens}, associated with the following prior,
taken from \citet{RichGreen}, which is symmetric with respect to the
components $k=1,\ldots,K$:
\begin{eqnarray*}
  \mu_k & \sim & \norm(m,\kappa^{-1}),\\
  \lambda_k & \sim & \gam(\alpha,\beta),\\
  \beta & \sim & \gam(g,h), \\
  (q_1,\ldots,q_{K-1}) & \sim & \dirich_K(1,\ldots,1). 
\end{eqnarray*}
In our examples, we take $m=M$, $\kappa = 4/R^2$, $\alpha=2$, $g=0.2$,
$h=100g/\alpha R^2$, where $R$ and $M$ are respectively
the range and the mean of the observed data, as in
\cite{jasra2005markov}. The posterior density reads
\begin{eqnarray}
 p(\theta|y) & = &  \frac 1 {Z_K} p(\theta) p(y|\theta) 
 \label{eq:posterior_density} \\
 & = &  \frac {\kappa^{K/2}g^h\beta^{K\alpha+g-1}}
{Z_K \Gamma(\alpha)^K\Gamma(g)(2\pi)^{\frac{n+K}{2}} }
\left(\prod_{k=1}^{K}\lambda_{k}\right)^{\alpha-1}
\exp\left\{ 
-\frac \kappa 2  \sum_{k=1}^K(\mu_k-M)^2
-\beta \left(h+\sum_{k=1}^K\lambda_k\right)\right\} \nonumber \\
& &  
\times \prod_{i=1}^n 
\left[\sum_{k=1}^K q_k \lambda_k^{1/2}
\exp\left\{ -\frac{\lambda_k}{2} (y_i-\mu_k)^2\right\}
\right]. \nonumber
\end{eqnarray}
In this expression, $\theta$ is the vector
\begin{equation}
\label{eq:deftheta}
\theta = (q_1,\ldots,q_{K-1},\mu_1,\ldots,\mu_K,\lambda_1,\ldots,\lambda_K,\beta)
\in \Omega= \mathcal{S}_K\times \R^{K}\times (\R_{+})^{K+1},
\end{equation}
the set 
\[
\mathcal{S}_{K}= \left\{ (q_1, \ldots, q_{K-1}) \in (\R_+)^{K-1}, \ \ \sum_{i=0}^{K-1} q_i \le 1 
\right \}
\] 
is the probability simplex, and 
\begin{equation}
  \label{eq:def_Z_K}
  Z_K = \int_\Omega p(\theta) p(y|\theta) \, d \theta
\end{equation}
is the normalizing constant (namely the marginal likelihood in~$y$), 
which depends on~$K$ and~$y$.

The sampling of the posterior distribution with density~\eqref{eq:posterior_density}
is the focus of the paper.

\subsection{Metastability in Statistical Physics} \label{sec:meta}

In this section, we draw a parallel between the problem of 
sampling~\eqref{eq:posterior_density} and the problem of sampling Boltzmann-Gibbs
probability measures that arise in computational Statistical Physics; 
see for instance~\citep{Balian}.  We take this opportunity to
introduce some of the concepts and terms of computational Statistical Physics that
are relevant in our context.

In computational Statistical Physics, one is often interested in
computing average thermodynamic properties of the system under
consideration. This requires sampling a Boltzmann-Gibbs (probability)
measure
\begin{equation}
  \label{eq:Gibbs_measure}
  \pi(\theta) = \frac 1 Z \exp\left\{-V(\theta)\right\}, 
\qquad Z = \int_\Omega \exp\left\{-V(\theta)\right\} \, d\theta,
\end{equation}
where $\theta\in\Omega\subset\R^d$, 
and $V$ is the \emph{potential} of the system, assumed to be such that $Z < \infty$. 
From now on, the term potential refers to the logarithm 
of a given (possibly unnormalized) probability density: \emph{e.g.} $V(\theta)=-\log\{p(\theta)
p(y|\theta)\}$ for the posterior density~\eqref{eq:posterior_density}. 

Probability densities such as the posterior
density~\eqref{eq:posterior_density} for mixture models, or the
Boltzmann-Gibbs density~\eqref{eq:Gibbs_measure} for systems in Statistical Physics,
are often multimodal.  Standard sampling strategies, for instance the 
Hastings-Metropolis algorithm~\citep{MRRTT53,Hastings70} 
generate samples which typically remain stuck in a 
small region around a local maximum (also called local mode)
of the sampled distribution (or, equivalently, a local minimum of the potential~$V$).
The sequences of samples generated by these algorithms are said to be \emph{metastable}.

\begin{figure}[htbp]
  \center 
 \includegraphics[width=7.1cm,height=5.cm]{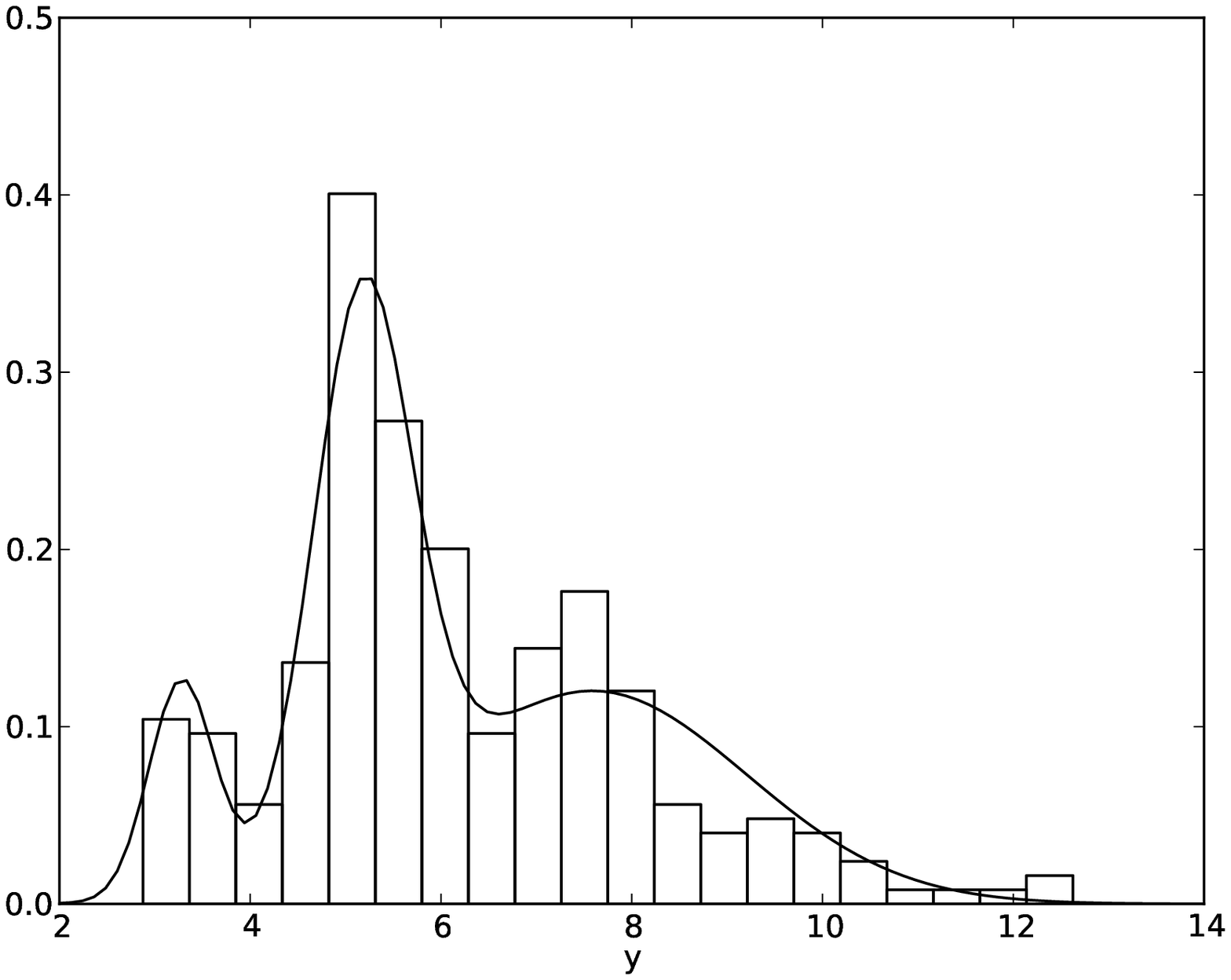}
 \includegraphics[width=7.3cm,height=5.cm]{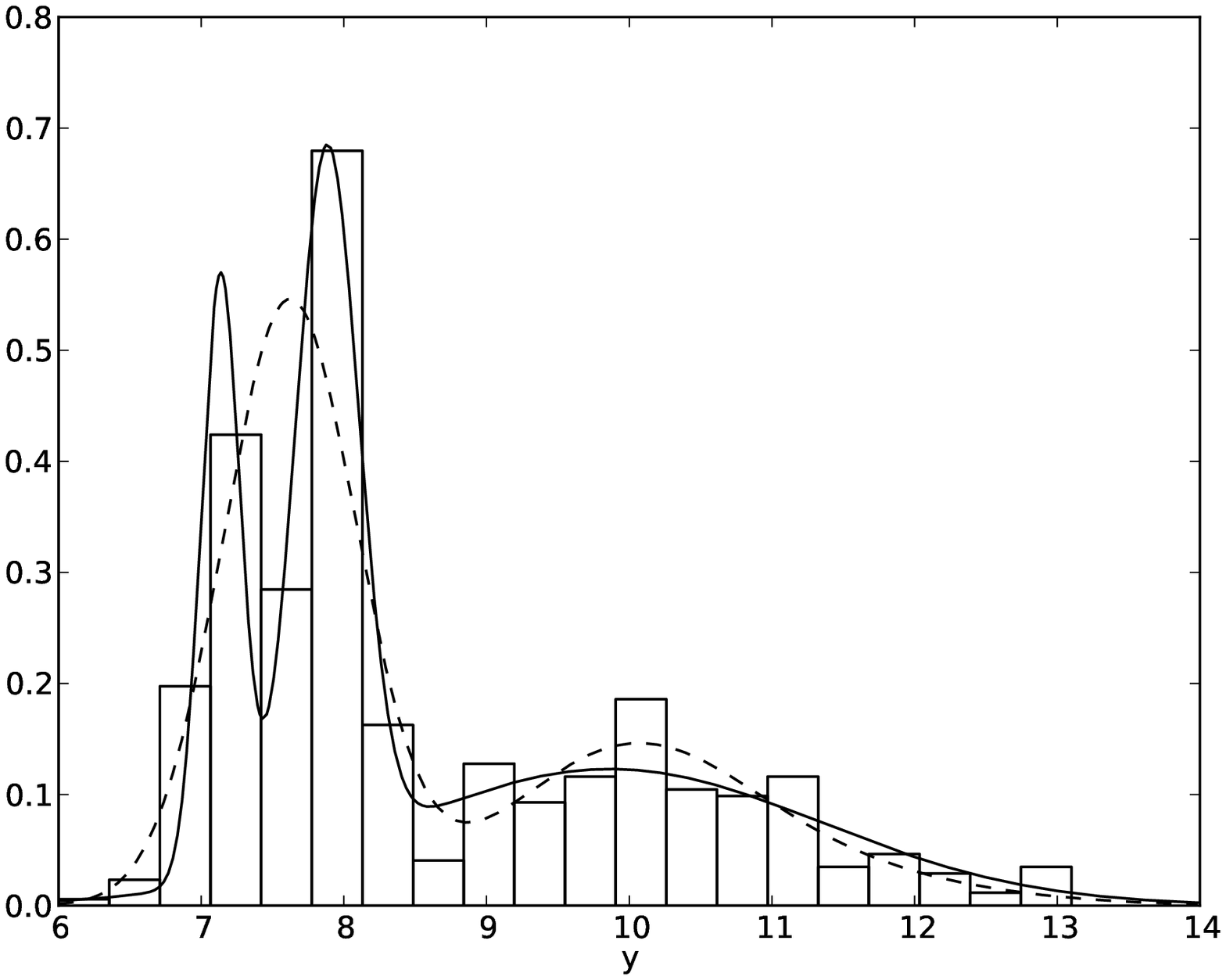} \\ 
  \includegraphics[width=7.3cm]{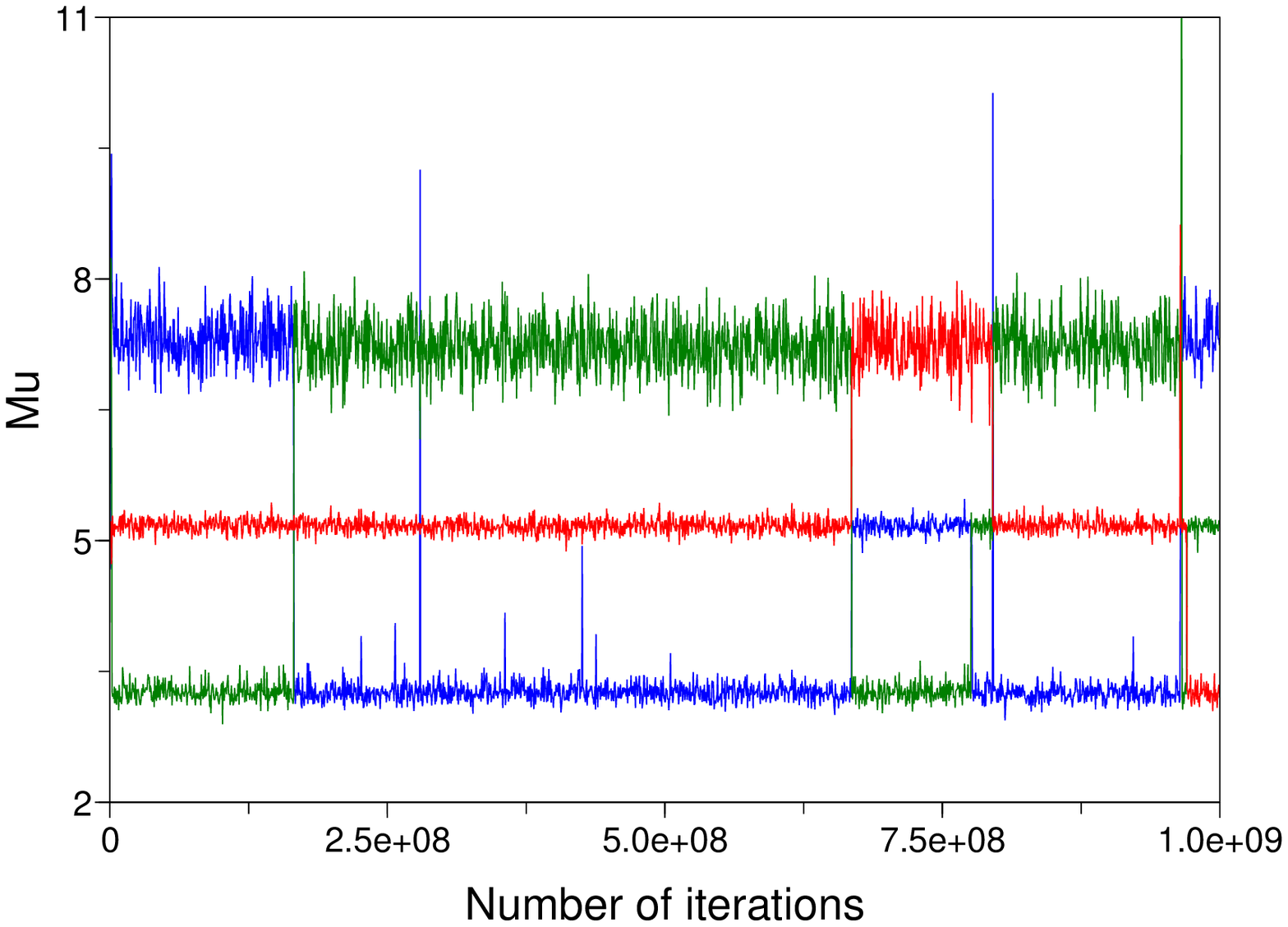} 
  \includegraphics[width=7.3cm]{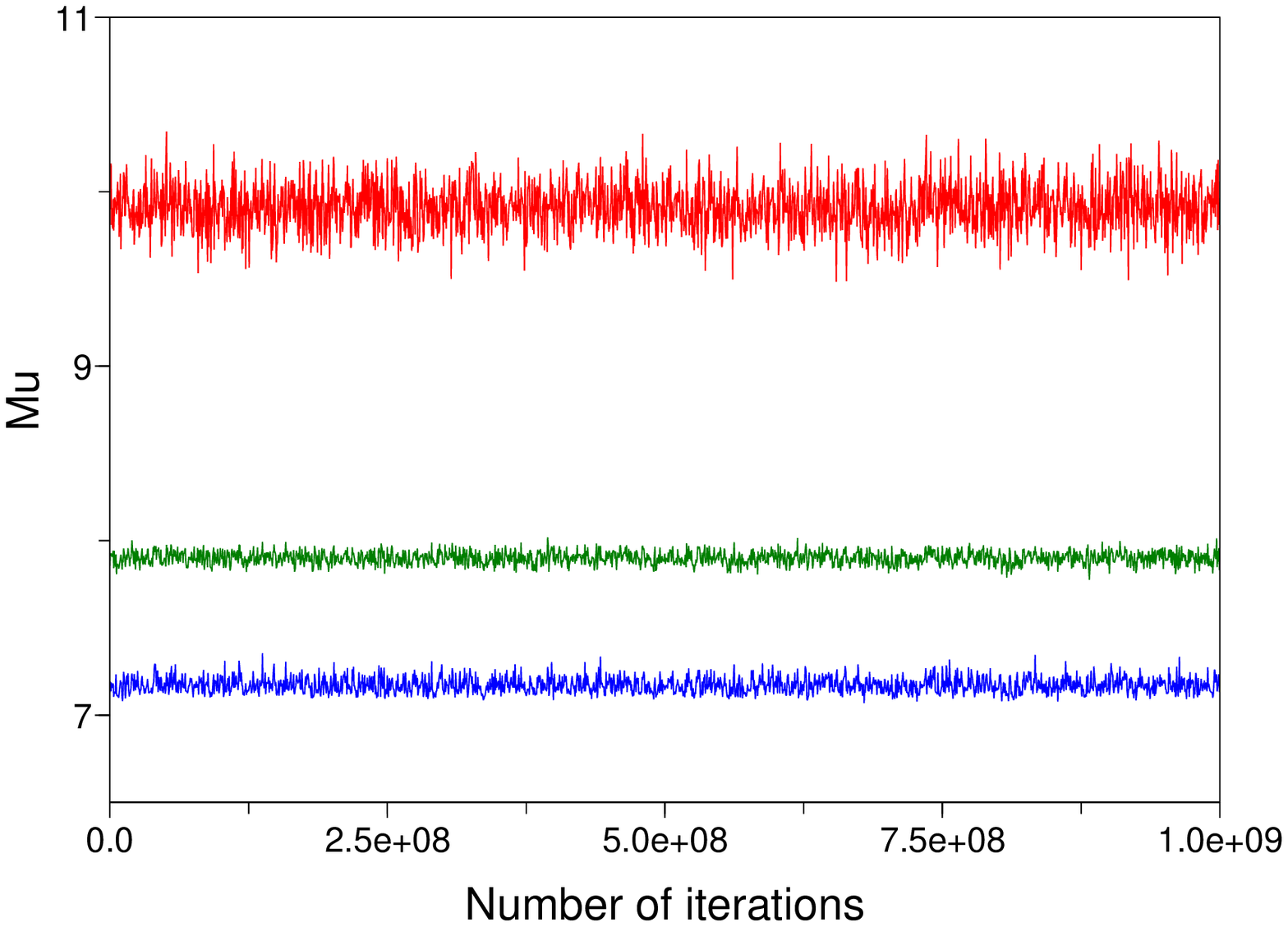}
  \caption{\label{fig:no_bias} Left (resp. right) hand side
    corresponds to Fishery (resp. Hidalgo stamps) dataset. Top row:
    histograms of the data and an estimated 3-component Gaussian
    mixture probability density function (solid line).  The dashed
    line corresponds to a local mode of the posterior density. Bottom
    row: random walk Hastings-Metropolis trajectories of $(\mu_1,\mu_2,\mu_3)$ as a function of
    the number of iterations. See Section~\ref{sec:examples} for more
    details on the data and the sampler. }
\end{figure}

Figure~\ref{fig:no_bias} illustrates this phenomenon, in the Bayesian
mixture context, for the posterior~\eqref{eq:posterior_density},
with $K=3$ components, and for two datasets (Fishery data, see
Section~\ref{sec:fishery}, and Hidalgo stamps data, see
Section~\ref{sec:hidalgo}). The posterior density  admits at
least $K!=6$ equivalent modes, but very few mode switchings (if any) are
observed in the MCMC trajectories. A simple Gaussian random walk
Hastings-Metropolis is used, but we obtained
the same type of plots (not shown) with Gibbs sampling.

The top right plot of Figure \ref{fig:no_bias} also represents a
Gaussian mixture probability density (in dashed line) which
corresponds to a negligible (in terms of posterior probability) 
local mode of the posterior density for
the Hidalgo dataset. In numerical tests not reported here,
when this local mode is used as a starting point,
the Hastings-Metropolis sampler used needs about $T=10^5$ iterations
to leave the attraction of this meaningless mode. It is easy to
make this problem worse by slightly changing the data. For instance,
$T$ is increased to $10^7$ by adding~2 to the three largest $y_i$'s
(while this local mode  remained of very small posterior probability). This
local mode illustrates the typical ``genuine multimodality'' of
mixture posteriors mentioned in the introduction --
multimodality which cannot be cured by label permutation.


\section{Free-energy biased sampling}
\label{sec:free_energy_biased_sampling}

Consider a generic probability density $\pi(\theta)$, with $\theta\in\Omega\subset\R^d$ 
as defined in \eqref{eq:Gibbs_measure}.
The principle of free energy biased sampling 
(described more precisely in Section~\ref{sec:principle}) 
is to change the original density $\pi$ to the biased density:
\[
\pi_A(\theta) \propto \pi(\theta) \exp\left\{(A\circ\xi)(\theta)\right\},
\]
where $\xi$ is some real-valued function $$\xi \, : \Omega \to [z_{\rm
  min},z_{\rm max}]$$ where $[z_{\rm min},z_{\rm max}] \subset
\mathbb{R}$ is a bounded interval, $(A\circ\xi)(\theta) = A(\xi(\theta))$, and
the so-called free energy~$A$ (see
definition~\eqref{eq:free_energy} below) is such that
the distribution of $\xi(\theta)$ when $\theta$ is distributed according to $\pi_A$ is
uniform over the interval $[z_{\min},z_{\max}]$. By sampling $\pi_A(\theta)\,d\theta$
rather than $\pi(\theta)\,d\theta$, the aim is to remove the metastability in
the direction of~$\xi$.  Averages with respect to the original
distribution of interest $\pi(\theta)\, d\theta$ are finally obtained by
standard importance sampling (see Section~\ref{sec:reweighting}). 
We assume first that $\xi(\theta)$ takes values in a bounded
interval $[z_{\rm min},z_{\rm max}]$,
(think of $\xi(\theta)=q_1$ and $[z_{\rm min},z_{\rm max}]=[0,1]$ for the mixture posterior
distribution case), 
and postpone the discussion of how to treat
reaction coordinates with values in an unbounded domain to
Section~\ref{sec:truncation}.

An important part of the algorithm is to compute (an approximation of) the 
free energy~$A$. There are many ways to this end. We describe 
a class of methods which are very efficient in the field of 
computational Statistical Physics  (see Section~\ref{sec:adaptive_strategy})
 and which, to our knowledge, have not been used so far in Statistics.

\subsection{Principle of the method}
\label{sec:principle}

Consider the conditional probability measures associated with~$\xi$:
\[
\pi^\xi(d\theta  \mid \xi(\theta)=z) = \frac
{ \dps \exp\{-V(\theta)\} \, \delta_{\xi(\theta)-z}(d\theta) } 
{ \dps \int_{\Sigma(z)} \exp\{-V(\theta')\} \, \delta_{\xi(\theta')-z}(d\theta') },
\]
where $\delta_{\xi(\theta)-z}(d\theta)$ is a measure with support 
\[
\Sigma(z) = \Big\{ \theta \in \Omega \ \Big| \ \xi(\theta) = z \Big\},
\]
defined by the formula: for all smooth test functions~$\varphi$ and $\psi$,
\[
\int_\Omega \psi(\xi(\theta)) \varphi(\theta) \, d\theta = \int \psi(z) \int_{\Sigma(z)} \varphi(\theta) \, 
\delta_{\xi(\theta)-z}(d\theta) \, dz.
\]
The main assumption underlying free-energy biased methods is that the
function $\xi$ is chosen so that the sampling of $\pi^\xi(d\theta \mid
\xi(\theta) = z)$ is significantly ``easier'' than the sampling of
$\pi(\theta)\, d\theta$, at least for some values of $z$. In other words,
$\pi^\xi(d\theta \mid \xi(\theta) = z)$ should be much less multimodal than
$\pi(\theta) \, d\theta$, at least for some values of $z$, see the discussion in
Section~\ref{sec:react-coord-mixt} below.

To give a concrete example, consider the choice $\xi(\theta)= q_1$.
In this case, $\pi^\xi(d\theta  \mid \xi(\theta)=z)$ is the conditional posterior distribution 
of all variables except $q_1$, conditionally on $q_1 = z$.

The function $\xi$ is called a reaction coordinate in the physics literature, because of its
physical interpretation: this function $\xi$ parameterizes the progress of some chemical event
at a coarser scale (chemical reaction or change of conformation for example). 
Given the trajectory $\{ \theta_t \}_{t \geq 0}$ of a Markov chain,
$\xi(\theta_t)$ is typically a metastable trajectory, 
and varies on timescales much larger than the typical microscopic fluctuations of the
system around its metastable configurations.
Of course, in our Bayesian mixture context, this
physical interpretation is not very useful. For the moment, we stick
with the more generic (and informal) understanding mentioned at the
beginning of this paragraph, \emph{i.e.} the sampling of $\pi^\xi(d\theta \mid
\xi(\theta) = z)$ should be ``easier'' than the sampling of $\pi(\theta)\,
d\theta$. We defer the important discussion on how to interpret and choose
this ``reaction coordinate'' in our specific context to
Section~\ref{sec:react-coord-mixt}. We also refer the readers
to~\cite{LRS08,LM10} for a precise quantification of this concept in a
functional analysis framework. Finally, although we consider only
one-dimensional reaction coordinates in this paper, we mention that
extensions to reaction coordinates with values in $\mathbb{R}^m$ with
$m\geq 2$ are possible~\citep{LRS10,CL10}. Some algorithms allowing to
switch between different reaction coordinates have also been
developed~\citep{PL07}.

The \emph{free energy} $A(z)$ is defined as
\begin{equation}
  \label{eq:free_energy}
  \exp\left\{-A(z)\right\} 
  = \int_{\Sigma(z)} \exp \left\{  -V(\theta) \right\} \, \delta_{\xi(\theta)-z}(d\theta),
\end{equation}
see for instance Section~5.6 in~\cite{Balian}. In other words, the
free energy is minus the marginal log-density of the reaction
coordinate. 
As above, let us again consider the simple example when the
reaction coordinate is $\xi(\theta) = q_1$.
Then, the free-energy is simply, up to an additive constant, equal to
\[
A(q_1) = - \log \left ( \int_{\mathcal{S}_{K-1}(q_1) \times 
\R^K \times (\R_+)^{K+1}} \exp\left\{-V(\theta)\right\} \, dq_2
\ldots dq_{K-1} \, d\mu_1 \dots d\mu_K \, d\lambda_1 \dots d\lambda_K \, d\beta \right ),
\]
where 
\[
\mathcal{S}_{K-1}(q_1) = \left\{ (q_2,\dots,q_{K-1}) \in (\R_+)^{K-2}, 
\ \ \sum_{i=2}^{K-1} q_i = 1 - q_1\right\}.
\]
In words, the free energy is in this case 
the opposite of the log-marginal density of $q_1$.

The free energy can be used to bias the target density $\pi$ as
follows: 
\[
\pi_{A}(\theta) = \frac{1}{Z_A} \exp\left\{-V(\theta)+(A\circ\xi)(\theta)\right\}.
\]
We refer to densities of this form as \emph{free energy-biased}
densities. The essential property of $\pi_A(\theta)$ is that, by
construction, the corresponding marginal distribution of~$\xi$ is
uniform on the interval $[z_{\min},z_{\max}]$.  A sampler targeting
$\pi_A(\theta)\,d\theta$ is thus much less likely to be \emph{metastable}
(namely to get stuck around a local minimum of the density) than a
sampler targeting $\pi(\theta)\,d\theta$, because (i) the former sampler should
move freely along the direction $\xi(\theta)$ defined by the reaction
coordinate, since the marginal distribution of $\xi(\theta)$ is uniform,
and (ii) we have assumed that the reaction coordinate $\xi(\theta)$ is such
the conditional probability distributions $\pi_A^\xi(d\theta \mid \xi(\theta)=z)
= \pi^\xi(d\theta \mid \xi(\theta)=z)$ are easy to sample, at least for some
values of $z$ (namely that they do not have very separated modes).

Therefore, free energy-based methods aim at sampling $\pi_A(\theta)$, in order to move
freely across the sampling space. Then, $\pi$ is eventually recovered through 
an importance sampling step, from  $\pi_A$ to  $\pi$: 
 for any test function~$\varphi$,
 \begin{equation}
   \label{eq:reweighting_theoretical}
   \mathbb{E}^\pi(\varphi) = \int_\Omega \varphi(\theta) \, \pi(\theta) \, d\theta = 
  \frac{\dps \int_\Omega \varphi(\theta) \exp\left\{-A\circ\xi(\theta)\right\}\, \pi_A(\theta) \, d\theta }{\dps  \int_\Omega \exp\left\{-A\circ\xi(\theta)\right\}\, \pi_A(\theta) \, d\theta}
   = \frac{\dps \mathbb{E}^{\pi_A}\Big(\varphi \exp\left\{-A\circ\xi\right\} \Big)}{\dps \mathbb{E}^{\pi_A}\Big(\exp\left\{-A\circ\xi\right\} \Big)}.
 \end{equation}
 We refer to Section~\ref{sec:reweighting} for further
 precisions. Note that for~\eqref{eq:reweighting_theoretical} to hold, $A$ only needs to be defined up to an additive constant.


\subsection{Computing the free energy by adaptive methods} 
\label{sec:adaptive_strategy}

In most cases, the free energy~$A$ defined in~\eqref{eq:free_energy} 
does not admit a closed-form expression, and must be estimated.
There are nowadays many techniques to this end, with various degrees of efficiencies
and conceptual complexities. 
We present in this section some powerful algorithms, 
namely adaptive biasing methods, which are
not so well known in the statistical literature. Of course, 
any other standard method such as thermodynamic integration could be used
(see the book~\citep{LRS10}
for a precise presentation of standard methods for free energy computations
in the framework of computational Statistical Physics,
as well as~\cite{gelman-meng-98} for a review from the viewpoint of Statistics). 

\subsubsection{General structure of adaptive methods}

In adaptive biasing Markov chain Monte Carlo methods, a time-varying
biasing potential $A_t(z)$ is considered.  The biasing potential $A_t$
is sequentially updated in order to converge to the free energy $A$ in
the limit. As already mentioned in the introduction, the term
``adaptive'' refers in this paper to the dynamic adaptation of the
targeted probability measure, and not of the parameters of a Markov
kernel used in the simulations. Specifically, at iteration~$t$, the
time-varying targeted density is
\begin{equation}
  \label{eq:pibt}
  \pi_{A_t}(\theta) = \frac 1 { Z_{A_t}}
  \exp\left\{-V(\theta)+(A_t\circ\xi)(\theta)\right\}. 
\end{equation}
An adaptive MCMC algorithm simulates
a non-homogeneous Markov chain $(\theta_t)$, $t=1,2,\ldots$, using the
two following steps at iteration~$t$:
\begin{enumerate}[\quad (1)]
\item a MCMC move according to the current target $\pi_{A_t}$ defined 
  in~\eqref{eq:pibt}, 
  \[
  \theta_{t} \sim K_t(\theta_{t-1},\cdot),
  \]
  where $K_t$ is a Markov kernel leaving $\pi_{A_t}$ invariant;
\item the update of the bias to $A_{t+1}$, using a trajectory average,
  see Section~\ref{sec:update} below. 
\end{enumerate}
The first step may be done using a Hastings-Metropolis kernel for instance,
see Figure~\ref{algo:MC_ABF}. 

Before explaining the second step, let us mention how the discretization 
of the reaction coordinate values for the biasing potential $A_t$ is done in practice.
A simple strategy, which we adopt in this
paper, is to use predefined bins, and approximate the biasing
potential $A_t$ or its derivative $A_t'$ (with respect to $z$) by
piecewise constant functions.  Specifically, we consider $N_z$ bins of
equal sizes $\Delta z$,
\[
[z_{\rm min},z_{\rm max}] =
\bigcup_{i=0}^{N_z-1}[z_i,z_{i+1}], \quad z_i = z_{\rm min} + i \Delta
z, \quad \Delta z = \frac{z_{\rm max}-z_{\rm min}}{N_z}.
\]
Other discretizations may of course be used, but this is not the focus of this paper.

\subsubsection{Strategies for updating the bias}\label{sec:update}

Recall that the bottom line of adaptive methods is that $A_t$ should converge
to~$A$. Adaptive biasing methods can be classified into two categories,
depending on whether it is the free energy $A_t(z)$, or its derivative
$A'_t(z)$ with respect to $z$, which is updated. 
Instances of the first
strategy, called adaptive biasing potential (ABP) methods, include
nonequilibrium metadynamics~\citep{BLP06,RLLgMP06}, the Wang-Landau
algorithm~\citep{WL01PRL,WL01} and Self-Healing Umbrella
Sampling~\citep{marsili-barducci-chelli-procacci-schettino-06}. The
adaptive biasing force (ABF) methodology \citep{DP01,HC04,LRS07},
which is the main adaptive method used in this paper, is an instance of the second. 
From now on, we focus on two
particular strategies, one belonging to the ABP class, and another to
ABF class. 

The ABP strategy we choose is based
on~\cite{marsili-barducci-chelli-procacci-schettino-06}.  In
particular, we do not use the Wang-Landau algorithm, which is, to our
knowledge, the only ABP method discussed before in the Statistical
literature; see \emph{e.g.} \cite{atchade-liu-04}, \cite{Liang05} and
\cite{Liang2010trajectory}.  Indeed, a delicate point with the
Wang-Landau approach is how to choose the vanishing rate of the ``gain
factor''; see \emph{e.g.} \citet{Liang05}.  On the other hand, Self-Healing
Umbrella Sampling does not involve such an additional parameter to be
tuned. It consists in updating $A_t$ as follows. The biasing potential
for $z \in (z_i,z_{i+1})$ is initially set to $\exp\{-A_0(z)\} =
1/N_z$ (for all $i \in \{0,\ldots,N_z-1\}$), and then updated for all
$i \in \{0,\ldots,N_z-1\}$ and for $t \geq 1$ as
\begin{equation}
  \label{eq:space_discretization_ABP}
  \forall z \in (z_i,z_{i+1}), 
  \qquad
   \exp\{-A_t(z)\} =
   \frac{1}{Z_t} \left( 1 + \sum_{j=1}^{t-1} \ind{z_i \leq \xi(\theta_j) < z_{i+1}} 
   \exp \big[ -A_{j}\circ\xi(\theta_j) \big] \right),
\end{equation}
the normalization factor $Z_t$ being such that 
\[
\Delta z \, \sum_{i=0}^{N_z-1} \exp\left\{-A_t\left(\frac{z_{i}+z_{i+1}}{2}\right)\right\} = 1.
\]
The method may be understood as follows: If $\theta_j$ was indeed
distributed according to $\pi_{A_j}$ at all times~$j$, then the weight
$ \exp \big[-A_{j}(\xi(\theta_j)) \big]$, proportional to
$\pi(\theta_j)/\pi_{A_j}(\theta_j)$, would correct for the bias introduced at
iteration~$j$, and $\exp\{-A_t(z)\}$ would indeed be an estimator of
the probability that $\xi(\theta)\in(z_i,z_{i+1})$ when $\theta$ is distributed
according to the \emph{unbiased} density $\pi$, 
that is $\exp \{-A(z) \}$. Of course, since $A_t$
is varying in time, it is not exactly true that $\theta_j$ is distributed
according to $\pi_{A_j}$, but this reasoning yields at least an
intuition on the way the method is built.  It is easy to check that,
provided the method converges, the only possible limit for the biasing
potential $A_t$ is the free energy~$A$ (up to the discretization error introduced by 
the binning of the reaction coordinate values).
Generalizations of the update equation~\eqref{eq:space_discretization_ABP}
leading to higher efficiencies have recently been proposed in~\citep{dickson-09}.

Alternatively, the ABF strategy~\citep{DP01,HC04,LRS07} is based on the following formula
for the derivative of~$A$ (called the \emph{mean force}):
\begin{equation} 
  \label{eq:mean-force}
  A'(z) = F(z) = \E^{\pi} \Big(f(\theta) \, \Big | \, \xi(\theta) = z \Big),
 \end{equation}
where $f$ admits an analytic expression in terms of $\xi$ and $V$:
\begin{equation}
  \label{eq:f}
  f = \frac{\nabla V \cdot \nabla \xi}{|\nabla \xi|^2} -
  {\rm div}\left( \frac{\nabla \xi}{|\nabla \xi|^2}\right),
\end{equation}
where $\nabla$ is the gradient operator, and ${\rm div}$ is the
divergence operator.  
As shown in~\citep{LRS10}, 
formulas~\eqref{eq:mean-force}-\eqref{eq:f} may
be derived from the definition~\eqref{eq:free_energy} of the free
energy using the co-area
formula~\citep{evans-gariepy-92,ambrosio-fusco-pallara-00}.  In
the simple case $\xi(\theta) = q_1$, it is easy to prove
that~\eqref{eq:mean-force}-\eqref{eq:f} hold, and that $f = \partial
V/\partial q_1$.  As mentioned above, the conditional measure of $\pi$
with respect to $\xi(\theta)=z$ is the same as the conditional measure of
$\pi_{A_t}$ with respect to $\xi(\theta)=z$. Thus, a natural ABF updating
strategy is to compute at iteration $t$ the following approximation of
the mean force: for all $i \in \{0,\ldots,N_z-1\}$ and for $t \geq 1$,
\begin{equation}
  \label{eq:space_discretization_ABF}
  \forall z \in (z_i,z_{i+1}), 
  \qquad
  F_t(z)
  = \frac{\dps \sum_{j=1}^{t-1} f(\theta_j) \, \ind{z_i \leq \xi(\theta_j) \leq z_{i+1}}}
  {\dps \sum_{j=1}^{t-1} \ind{z_i \leq \xi(\theta_j) \leq z_{i+1}}}.
\end{equation}
From this approximation of $F$, an approximation $A_t$ of the free
energy~$A$ can be recovered by integrating~$F_t(z)$ in $z$. The consistency of
the method may be understood as follows: If $\theta_j$ was distributed
according to $\pi_{A_j}$ at all times~$j$, then we would have $F_t =
F$ and hence $A_t = A$ (up to an additive constant).  Besides, as
above, it can be shown that, provided the method converges, the
biasing potential $A_t$ converges to (a discretized version of) the
free energy~$A$, up to an additive constant.

The interest of ABP compared to ABF is that it does not require
computing $f$ given by~\eqref{eq:f}, which may be cumbersome for some~$\xi$ such as 
$\xi = V$ (minus the log posterior density).
On the other hand, it is observed that the ABF method yields very good
results since the derivative of the free energy is approximated, so that after
integration, the adaptive biasing potential is smoother in $z$ for ABF
than for ABP. In the following, we use the ABF method, except when we
consider as a reaction coordinate the potential $V$. In this case, the ABP method is used.

The convergence of the adaptive biasing force method (for a slightly different dynamics), has been
studied in~\citep{LRS08}, and its associated discretization using many
replica of the simulated Markov chain has been considered
in~\cite{jourdain-lelievre-roux-08}. For refinements concerning the
implementation of such a strategy, we refer to~\citet{LRS07}.

\subsubsection{Practical implementation of adaptive algorithms}

To summarize the method, we give in
Figure~\ref{algo:MC_ABF} the details of the ABF algorithm. A similar
algorithm is used in the ABP case. In practice, we stop the algorithm
when the bias $A_t$ is no longer significantly modified from $t=n$ to $t=n+N_{\rm cvg}$,
where $N_{\rm cvg}$ is a fixed number of iterations between two convergence checks.
See Section~\ref{sec:cv_adaptatif} for an illustration of this strategy.
 
The so-obtained bias is then considered as a good approximation of the
free energy, and used in the subsequent importance sampling step, as
explained in Section~\ref{sec:reweighting}. 
A perfect convergence is not required, in the
sense that the bias does not need to be estimated very accurately, as it is 
removed in the final importance sampling step.
Moreover, note that the biasing potential~$A_t$ needs to be computed only
up to an additive constant which does not play any role in 
the overall procedure.

\begin{figure}[htbp]
  \caption{ \label{algo:MC_ABF}  The Markov chain Monte-Carlo adaptive biasing
      force algorithm.}
  \bookbox{
    \begin{algo}
      Consider a reaction coordinate $\xi$. Starting from some
      initial configuration~$\theta_0$ and the biasing potential $A_0 = 0$, iterate on $t \geq 1$:
      \begin{enumerate}[\quad (a)]
      \item Propose a move from $\theta_{t-1}$ to $\theta'_{t}$ according to the
        transition kernel $\mathscr{T}(\theta_{t-1},\theta'_{t}) \, d\theta'_{t}$;
      \item Compute the acceptance rate
        \[
        \alpha_t = \min \left(\frac{\pi_{A_{t}}(\theta'_{t}) \,
          \mathscr{T}(\theta'_{t},\theta_{t-1})} {\pi_{A_{t}}(\theta_{t-1}) \,
          \mathscr{T}(\theta_{t-1},\theta'_{t})},\,1 \right),
        \]
        where the biased probability density $\pi_{A_t}$ is defined as
        \[
        \pi_{A_{t}}(\theta) \propto \pi(\theta) \exp\big[ A_{t}\circ{\xi}(\theta) \big];
        \]
      \item Draw a random variable $U_t$ uniformly distributed in
        $[0,1]$ ($U_t \sim {\cal U}[0,1]$);
        \begin{enumerate}[\quad (i)]
        \item if $U_t \le \alpha_t$, accept the move and set
          $\theta_{t} = \theta'_{t}$;
        \item if $U_t > \alpha_t$, reject the move and set $\theta_{t}
          = \theta_{t-1}$.
        \end{enumerate}
      \item Following~\eqref{eq:space_discretization_ABF}, 
        update the biasing force, hence the biasing potential~$A_{t+1}$.
      \item Go to Step~(a).
      \end{enumerate}
    \end{algo}
  }
\end{figure}

\subsection{Reweighting free-energy biased simulations}
\label{sec:reweighting}

Upon stabilization of the adaptive algorithm at iteration $T$, an
estimate $\widehat A  = A_T$ of the biasing potential $A$ is obtained, from which
one defines the biased density 
\begin{equation}
  \label{eq:biased_posterior}
  \widetilde{\pi}(\theta) = \pi_{\widehat A}(\theta) =  \frac 1 {\widetilde Z} \, \pi(\theta)  
  \exp \left \{ \widehat {A}\circ\xi(\theta) \right \}. 
\end{equation}
To sample the true posterior $\pi$, we use the following simple
strategy. We simulate a standard MCMC algorithm, \emph{e.g.} a random
walk Hastings-Metropolis algorithm, targeted at the biased posterior
density $\widetilde \pi$, and then perform an importance sampling step from 
$\widetilde \pi$ to  $\pi$, based on the importance sampling weights: 
\begin{equation}
  \label{eq:wis}
  w(\theta) =  \exp\left\{-\widehat{A}\circ\xi(\theta)\right\} \propto \frac{\pi(\theta)}{\widetilde{\pi}(\theta)}.
\end{equation}
From the MCMC chain $(\theta_t)_{t \ge 1}$ targeted at $\widetilde \pi$, the expectation with respect to$~\pi$ 
of a test function $h$ can thus be estimated as (see~\eqref{eq:reweighting_theoretical}):
\begin{equation}
  \mathbb{E}^\pi(h) = \frac{\mathbb{E}^{\widetilde{\pi}}(h w)}{\mathbb{E}^{\widetilde{\pi}}(w)} 
  \simeq \frac {\dps \sum_{t=1}^{t_{\rm max}} h(\theta_t) w(\theta_t)}
         {\dps \sum_{t=1}^{t_{\rm max}} w(\theta_t) }, \label{eq:isestimator}
\end{equation}
where $t_{\rm max}$ is the number of iterations of the MCMC chain.

\subsection{Reaction coordinates with unbounded values}
\label{sec:truncation}

There are many cases when the reaction coordinate takes values in an
unbounded interval~$\mathscr{I}$. Here $\mathscr{I}$ should be
understood as the support of the distribution of the random variable
$\xi(\theta)$ when $\theta \sim \pi(\theta) \, d\theta$. One may think of
$\xi(\theta)=\mu_1$ as an example for the mixture posterior
distribution case (in which case $\mathscr{I}=\mathbb{R}$), and see
Section~\ref{sec:react-coord-mixt} for more examples.

It is not possible to apply the above procedure on the whole interval $\mathscr{I}$. Some truncation is required for at least two reasons. First, numerically, it would be difficult to discretize in space a function defined on an unbounded domain. Second (and more importantly) the use of the full free energy over $\mathscr{I}$ would lead to a density $\pi_A$ which is not integrable (since the uniform law over $\mathscr{I}$ is not well defined as a probability distribution).

We therefore resort to the following strategy. First, we choose some
truncation interval $[z_{\min},z_{\max}]$. Then, in the adaptive MCMC
algorithm (which calculates the free energy), we reject any point $\theta$
such that $\xi(\theta)$ fall outside of this interval. This is tantamount
to restricting the sampling space with the constraint $z_{\min}\leq
\xi(\theta)\leq z_{\max}$. In this way, one obtains an estimate $\widehat A(z)$
of the free energy $A(z)$, but only for $z\in[z_{\min},z_{\max}]$. 

When $\widehat A$ is obtained, we simply extend its definition outside
$z\in[z_{\min},z_{\max}]$ as follows:  $\widehat A(z)=\widehat A(z_{\min})$, for $z\leq
z_{\min}$, $\widehat A(z) = \widehat A (z_{\max})$, for $z\geq z_{\max}$.
Finally, we run a standard MCMC sampler targeting the 
distribution $\widetilde{\pi}=\pi_{\widehat A}$, as described in the previous
section. Note that the biased distribution $\widetilde \pi$ is
 defined over the whole parameter space $\Omega$ (in particular, 
 no additional rejection step is needed in the sampling of this distribution).

In practice, one should choose an interval $[z_{\min},z_{\max}]$ which is not too
large, but at the same time such that the probability (with respect to
$\pi$) of the event $z_{\min}\leq \xi(\theta)\leq z_{\max}$ is close to
one:
\begin{equation}
  \label{eq:zmin_zmax}
\frac{\dps \int_{z_{\rm min}}^{z_{\rm max}} \, \exp(-A(z)) \, dz}
     {\dps \int_{\mathscr{I}} \, \exp(-A(z)) \, dz} \simeq 1,
\end{equation}
so that $\mathscr{I} \setminus [z_{\rm min},z_{\rm max}]$ is barely visited (see also Section~\ref{sec:efficiency}). This is one of the practical difficulties that we shall discuss
in the next section.

\section{Bayesian inference from free energy biased dynamics}
\label{sec:abf-bayes-mixt}

In this section, we explain how to perform Bayesian inference for the
univariate Gaussian mixture model described Section
\ref{sec:post-distr}, that is, how to compute quantities such as
posterior expectations and ratios of marginal likelihoods (equal to
ratios of normalizing constants~$Z_{K}$ defined
in~\eqref{eq:def_Z_K}), using the free energy associated to a given
reaction coordinate to build an importance function.  The Gaussian mixture
model corresponds, in the notation of
Section~\ref{sec:free_energy_biased_sampling}, to 
$\pi(\theta) = p(\theta|y)\propto p(\theta)p(y|\theta)$, hence
$V(\theta) = -\log \{ p(\theta) p(y|\theta) \}$. For a given reaction
coordinate, and a given estimate $\widehat{A}$ of the free energy $A$,
the free-energy biased probability distribution is
\[
\widetilde{p}(\theta|y) \propto p(\theta|y) / w(\theta) 
\propto p(\theta|y) \exp\left(\widehat{A} \circ \xi\right),
\] 
where $w$ is defined by~\eqref{eq:wis}.

We start by listing the criteria we use to assess the quality 
of the importance sampling procedure in Section~\ref{sec:criteria}.
As mentioned in the introduction, the strategy to sample the posterior
distribution~\eqref{eq:posterior_density} consists of three steps:
choosing a reaction coordinate, computing (an approximation of) the
free energy associated to this reaction coordinate, and using the free
energy to build an importance sampling proposal distribution according 
to~\eqref{eq:biased_posterior}. 
The previous section was
devoted to the second and third steps. We discuss the first
step in Section~\ref{sec:react-coord-mixt} for the mixture model at
hand. Section~\ref{sec:comp-norm-const} presents an extension of the
method to the computation of the ratio of normalizing constants
associated to different values of the number of components~$K$, in order to
perform model choices between models corresponding to different
number of components. Note that we discuss in this section the
theoretical efficiency of the whole approach. These
discussions are supported by numerical experiments in
Section~\ref{sec:examples}.
 
\subsection{Criteria for choosing the reaction coordinate}
\label{sec:criteria}

\subsubsection{General criteria}

We consider the following criteria for evaluating the practical
efficiency of the whole procedure, for a given choice of the reaction
coordinate $\xi$:
\begin{enumerate}[(a)]
\item In the execution of the (either ABF or ABP) adaptive algorithm,
  how fast does the approximate free energy $A_t$ converge to its
  limit $A$?
\item How efficient is the
  importance sampling step from the biased distribution to the originally targeted
  posterior distribution? Actually, this criterion is twofold: 
  \begin{enumerate}[(b1)]
  \item How
    efficient is the MCMC sampling of the biased density $\widetilde{p}(\theta|y)$? 
  \item How
    representative are the points simulated from the biased distribution with respect to the target
    posterior distribution? (\emph{i.e.} how many of these points are assigned 
    non-negligible importance weights?)
  \end{enumerate}  
\item A more practical criterion is (in the case of a reaction
  coordinate with values in an unbounded domain): How difficult is it
  to determine, a priori, an interval $[z_{\min},z_{\max}]$ for the
  reaction coordinate values, which ensures good performance with
  respect to Criteria~(a) and (b) and which
  satisfies~\eqref{eq:zmin_zmax} ?
\end{enumerate}
Criterion~(b2) is discussed in the next section.  Criteria~(a)
and~(b1) can actually be shown to be closely related, at least for
some family of adaptive methods, see~\cite{LRS08}. Roughly speaking,
an adaptive algorithm yields quickly an estimate of the free energy,
if and only if the free energy is indeed a good biasing potential, in
the sense that the dynamics driven by the biased potential converges
quickly to a limiting distribution. Theoretically and as mentioned
above, a sufficient condition for an efficient sampling is that the
conditional probability distributions $\pi^\xi(dx \mid \xi(x)=z)$ are
easy to sample, at least for some values of~$z$ (namely they do not
have very separated modes). We refer to \cite{LRS08,LM10} for precise
mathematical results.  

Numerically, to assess the convergence of
adaptive methods, we recommend the following two basic checks: (i)
that the output of the adaptive algorithm has explored the full range
$[z_{\min},z_{\max}]$ and has a distribution which is close to
uniform; and (ii) using the criterion mentioned in the introduction,
and specifically for mixture posterior distributions, that the
algorithm has visited the $K!$ symmetric replicates of any significant
local mode. The same convergence checks can be applied to the MCMC
dynamics targeted at $\widetilde{\pi}$.

\subsubsection{Efficiency of the importance sampling step}\label{sec:efficiency}

We  give here  a way to quantify Criterion~(b2).
To evaluate the performance of the importance sampling step, 
we compute the following efficiency factor 
\[
\mathrm{EF} = \frac 
{\dps \left(\sum_{t=1}^{T} w(\theta_t)\right)^2}
{\dps T \sum_{t=1}^{T} w(\theta_t)^2}
\]
where $w(\theta)$ is defined in~\eqref{eq:wis}, and where 
$\{ \theta_t \}_{t \ge 0}$ denotes the MCMC sample targeting 
 the biased posterior $\widetilde{p}(\theta|y)$, as described in Section \ref{sec:reweighting}.
The efficiency factor is the 
Effective Sample Size of~\cite{KongLiuWong} divided by
the number of sampled values.  This quantity lies in
$[0,1]$. It is close to one (resp. to zero)  when the random variable $w(\theta)$ has a
small (resp. a large) variance. Indeed, it is easy to check that
\[
\mathrm{EF}=\left(\frac{{\rm Var}_T(w)}{({\rm E}_T(w))^2}+1
\right)^{-1},
\qquad
{\rm Var}_T(w) = \frac{1}{T} \sum_{t=1}^{T} w(\theta_t)^2 - 
\Big[ {\rm E}_T(w)\Big]^2,
\]
where the latter quantity is the empirical variance of the sample $\{w(\theta_t)\}_{1 \le t \le T}$, 
and ${\rm E}_T(w)=  \sum_{t=1}^{T} w(\theta_t)/T$ its empirical average.

We now propose an estimate of the efficiency factor
in terms of the converged bias $\widehat A$ only, which may therefore be
computed \emph{before} the MCMC algorithm targeting the biased
posterior is run, 
and the importance sampling step is performed. This estimate
is based on the fact that, with respect to $\widetilde{p}(\theta|y)$, the
marginal distribution of $\xi$ is approximately uniform over $[z_{\min},z_{\max}]$.
For well chosen $z_{\rm min}$ and $z_{\rm max}$, $\xi(\theta_t)$ hardly 
visits values out of the interval $[z_{\rm min},z_{\rm max}]$ (see~\eqref{eq:zmin_zmax} above) and thus
\[
\frac{{\rm Var}_T(w)}{({\rm E}_T(w))^2} 
\simeq \frac{\displaystyle \frac{1}{z_{\rm max} - z_{\rm min}} 
\int_{z_{\rm min}}^{z_{\rm max}} \left(\exp\left\{-\widehat{A}(z)\right\} - 
\frac{1}{z_{\rm max} - z_{\rm min}}\int_{z_{\rm min}}^{z_{\rm max}} \exp\left\{-\widehat{A}\right\} 
\right)^2 \, dz}
{\displaystyle \left(\frac{1}{ z_{\max}-z_{\min}} \int_{z_{\rm min}}^{z_{\rm max}} 
\exp\left\{-\widehat{A}(z)\right\} \, dz \right)^2},
\]
which provides a justification for the following ``theoretical''
efficiency factor: 
\begin{equation}
  \label{eq:formula_EF}
  \mathrm{EF}_{\mathrm{theoretical}} = 
  \frac{\displaystyle \left( \int_{z_{\rm min}}^{z_{\rm max}}
      \exp\left\{-\widehat A(z)\right\} \, dz \right)^2}
       {\displaystyle (z_{\rm max} - z_{\rm min}) \int_{z_{\rm
             min}}^{z_{\rm max}} \exp\left\{-2\widehat A(z)\right\} \, dz}. 
\end{equation}
The agreement between theoretical and numerically computed efficiency
factors in our simulations is very good, see Tables~\ref{tab:fish_EF_RC}, 
\ref{tab:fish_EF} and~\ref{tab:hidalgo_EF_RC}
in Section \ref{sec:appl-fish-data}. Thus, the theoretical efficiency
factor allows for a quick check that the subsequent importance
sampling is reasonably efficient.

From the expression~\eqref{eq:formula_EF}, it is seen that the efficiency factor is
close to one when $A$ is close to a constant.  Thus, Criterion~(b2)
mentioned in the previous section is likely to be satisfied if the
free energy associated to $\xi$ has a small amplitude, \emph{i.e.} 
$\max A - \min A$ is as small as possible. 

\subsection{Practical choice of the reaction coordinate}
\label{sec:react-coord-mixt}

We now discuss the practical choice of the reaction coordinate
$\xi:\theta\rightarrow \R$ in the mixture posterior sampling context,
with respect to the criteria listed above.  We discuss successively
the following four possible choices: $\xi(\theta)=\mu_1$,
$\xi(\theta)=V(\theta)=-\log\{p(\theta)p(y|\theta)\}$,
$\xi(\theta)=q_1$ and $\xi(\theta)=\beta$.  This discussion is also
illustrated numerically in Section~\ref{sec:eff_samp_num}.

The requirement that the multimodality of the
target measure conditional on $\xi(\theta)=z$ is much less
noticeable than the multimodality of the original target measure
rules out the choice of $\mu_1$ as a good reaction
coordinate since, conditionally on~$\mu_1$, the posterior density still
has at least $(K-1)!$ modes, as the components $2$ to $K$ remain exchangeable. 
Numerical tests indeed support these considerations, see below.

A more natural reaction coordinate is minus the posterior log-density
$\xi(\theta)=-\log\{p(\theta)p(y|\theta)\}$, in the spirit of the original
Wang-Landau algorithm~\citep{WL01PRL,WL01}. Indeed, exploring regions
of low posterior density should help to escape more easily from local
modes. Unfortunately, determining a range
$[z_{\min},z_{\max}]$ of ``likely values'' (with respect to the
posterior distribution) for such functions of $\theta$ is not
straightforward; see Criterion~(c) above. Moreover,
since the posterior density is expected to be multimodal and difficult
to explore, there seems to be little point in performing MCMC pilot
runs in order to determine $[z_{\min},z_{\max}]$. A conservative
approach is to choose a very wide interval $[z_{\min},z_{\max}]$, but
this makes the subsequent importance sampling step quite
inefficient. In our simulations, we report satisfactory results for
this reaction coordinate, but with the caveat that our choice for
$[z_{\min},z_{\max}]$ was facilitated by our different simulation
exercises, based on several reaction coordinates. Another practical
difficulty we observed in one case is that the estimated bias is
quite inaccurate in the immediate vicinity of the posterior mode,
because the free energy tends to increase very sharply in this region,
see Section \ref{sec:hidalgo} for more
details. 

The choice $\xi(\theta)=q_1$ is satisfactory with respect to Criterion~(c): 
The range on which it can vary, namely $[0,1]$, is clearly
known. With respect to (a), this choice looks appealing as well, since
forcing $q_1$ to get close to~1 should empty the $K-1$ other
components, which then may swap more easily. 
Unfortunately, we
observe in some of our experiments that the dynamics biased
by the free-energy associated with this reaction coordinate 
is not very successful in terms of mode switchings, see 
Figures~\ref{fig:fish_bias_traj} and~\ref{fig:hidalgo_bias_traj}. 

Finally, $\xi(\theta)=\beta$ appears to be a good trade-off with
respect to our criteria, 
at least in the examples we consider below. Concerning the
determination of the interval $[z_{\rm min},z_{\rm max}]$ (Criterion~(c)), 
since~$\beta$ determines the order of magnitude of the component
variances $\sigma_k^2=\lambda_k^{-1}$, there should be high posterior
probability that $\beta$ is a small fraction of~$R^2$, where~$R$ is
the range of the data. For instance, we obtain satisfactory results
in all our experiments with
$[z_{\min},z_{\max}]=[R^2/2000,R^{2}/20]$. Concerning Criterion~(a), we
observe that the choice $\xi=\beta$ performs well (see the numerical results
below). 

We propose the following explanation. Since the $\lambda_k$'s have a
$\gam(\alpha,\beta)$ prior, large values of~$\beta$ penalize large
values for the component precisions~$\lambda_k$, or equivalently
penalize small values for the component standard deviations
$\sigma_k=\lambda_k^{-1/2}$. If $\beta$ is large enough, the Gaussian
components are forced to cover the complete range of the data, and
thus can switch easily.  This interesting phenomenon is illustrated by
Figure~\ref{fig:mulambda_condbeta}, see below for further precisions. 
In other words, a ``good'' reaction coordinate $\xi$ should be such that the conditional probability distributions $\pi^\xi(dx \mid \xi(x)=z)$ are less multimodal than $\pi$, {\em at least for some values of~$z$}. For a theoretical result supporting this interpretation, we refer to~\cite{LM10}.

\subsection{Computing normalizing constants and model choice}
\label{sec:comp-norm-const}

In this section, we discuss an extension of the method to perform model 
choice between models with different numbers of
components. The principle is to compute the normalizing constant $Z_K$ of the
posterior density for different values of $K$, see~\eqref{eq:def_Z_K}
for a definition of~$Z_K$. More precisely, 
it is sufficient to evaluate $Z_K/Z_{K-1}$ for a given range of $K$ (see
\cite[Chap. 7]{RobCas} for a review on Bayesian model choice).

We propose the following strategy. The estimation of $Z_K/Z_{K-1}$ can
be performed by first estimating $Z_K/\widetilde Z$, then estimating
$Z_{K-1}/\widetilde Z$, and finally dividing the two quantities.  A
simple estimator of $Z_K/\widetilde Z$ (where $\widetilde Z$ is the normalizing
constant in~\eqref{eq:biased_posterior}) is given by
\[
\widehat{I}_{K} =\frac{1} {T} \sum_{t=1}^{T} w(\theta_t), 
\]
where $\{ \theta_t \}_{t \ge 0}$ is a sample distributed according 
to the biased probability $\widetilde{p}(\theta|y)$ (with $K$ normal components).
This formula is based on the fact that 
the expectation of $w(\theta)=\exp\{-\widehat A\circ\xi(\theta)\}$ with
respect to $\widetilde{p}(\theta|y)$ is $Z_K/\widetilde Z$. 

Let $\theta_{-k}$ denote the parameter vector obtained by removing in $\theta$
the parameters  attached to  a given component $k$,  and replacing the
probabilities $q_l$ (for $l \neq k$) by $\widetilde{q}_{l}=q_l/(1-q_k)$. 
Let  $p(y|\theta_{-k})$ denote 
the likelihood of the model with $K-1$ components, and parameter
$\theta_{-k}$. Then the following quantity
\[ 
\widehat{I}_{K-1} = \frac{1}{K}\sum_{k=1}^K \widehat{I}_{K-1,k},
\qquad \widehat{I}_{K-1,k} =  \frac{1}{T}  \sum_{t=1}^{T} w_{-k}(\theta_t), 
\]
where $\{ \theta_t \}_{t \ge 0}$ is the same Markov chain as above, and
\begin{equation}
\label{eq:wmk}
  w_{-k}(\theta) = \frac{p(y|\theta_{-k})}{p(y|\theta)}
\exp\left\{ -\widehat{A}\circ\xi(\theta) \right\}, 
\end{equation}
is an estimator of $Z_{K-1}/\widetilde Z$. 

The estimators $\widehat{I}_K$ and $\widehat{I}_{K-1}$ are reminiscent
of the birth and death moves of the reversible jump algorithm of 
\citet{RichGreen}, where a new model is proposed by adding or
removing a component chosen at random. The difference is
that the biased posterior $\widetilde{p}(\theta|y)$ acts as an intermediate
state between the posterior with $K$ components, $p(\theta|y)$ and
the posterior with $K-1$ components (or more precisely, the posterior
with $K-1$ components times the prior of a $K$-th ``non-acting''
component, in order to match the dimensionality of both
$p(\theta|y)$ and $\widetilde{p}(\theta|y)$).  

In our numerical experiments, the estimator of $Z_K/Z_{K-1}$ obtained from this 
strategy performs well, see Section \ref{sec:examples} (in particular 
Table~\ref{tab:fish}). 


\section{Numerical examples} 
\label{sec:examples}

In our experiments and as explained above, we use the following
approach. First, we run an adaptive algorithm (ABF, except
for $\xi(\theta)=V(\theta)=-\log\{p(\theta)p(y|\theta)\}$, in which case we use ABP), for a
given choice of the reaction coordinate $\xi$, and a given interval
$[z_{\min},z_{\max}]$, until a converged bias $\widehat A$ is
obtained. Second, we run a MCMC algorithm, with $\widetilde{p}(\theta|y)$ given
by~\eqref{eq:biased_posterior} as invariant density. Third, we perform
an importance sampling step from $\widetilde{p}(\theta|y)$ to $p(\theta|y)$, the unbiased
posterior density. See the introduction of Section~\ref{sec:abf-bayes-mixt}
for the notation.

The quality of the biasing procedure is assessed using the criteria
mentioned in Sections~\ref{sec:criteria}. This consists in: (i)
checking that the reaction coordinate values are uniformly sampled
over $[z_{\min},z_{\max}]$, (ii) checking that the output is symmetric
with respect to labellings, and many switchings between the modes are
observed and (iii) computing the efficiency factor (a good indicator
being the estimator~\eqref{eq:formula_EF} defined in terms of
$\widehat{A}$).

In the first step of the method (approximation of the free energy using
adaptive algorithm), we deliberately use the
simplest exploration strategy, namely a Gaussian random walk
Hastings-Metropolis update, with small scales
(see below for the precise values). This is to illustrate
that the ability of adaptive algorithms to approximate the free energy
does not crucially depend on a finely tuned updating strategy. 

In the second step, we run a Hastings-Metropolis algorithm targeted at
the biased posterior, using Cauchy random walk proposals, and the
following scales: $\tau_{\mu}=R/1000$, $\tau_v=2/R^2$,
$\tau_{\beta}=2\times 10^{-5}\alpha R^2$, where $R$ is the range of
the data, which leads to acceptance rates between $10\%$ and $30\%$ in
all cases.

\subsection{A first example : the Fishery data}
\label{sec:fishery}

We first consider the Fishery data of \citet{TittBookMixtures} (see
also~\citet{Fru:book}), which consist of the lengths of 256 snappers,
and a Gaussian mixture model with $K=3$ components; see
Figure~\ref{fig:no_bias} for a histogram.

\subsubsection{Convergence of the adaptive algorithms}
\label{sec:cv_adaptatif}

In the adaptive algorithm, we use Gaussian random walk proposals with
scales $\tau_q = 5\times 10^{-4}$, $\tau_\mu = 0.025$, $\tau_v = 0.05$
and $\tau_\beta = 5\times 10^{-3}$. These parameters were also used to
produce the unbiased trajectory in Figure~\ref{fig:no_bias}. We
illustrate here the convergence process in the case
$\xi(\theta) = \beta$, using the ABF algorithm described in
Section~\ref{sec:adaptive_strategy}, with $z_{\rm min} = 0.05$,
$z_{\rm max} = 4.0$ and $\Delta z = 0.01$.

\begin{figure}[htbp]
  \center
  \includegraphics[width=7.3cm]{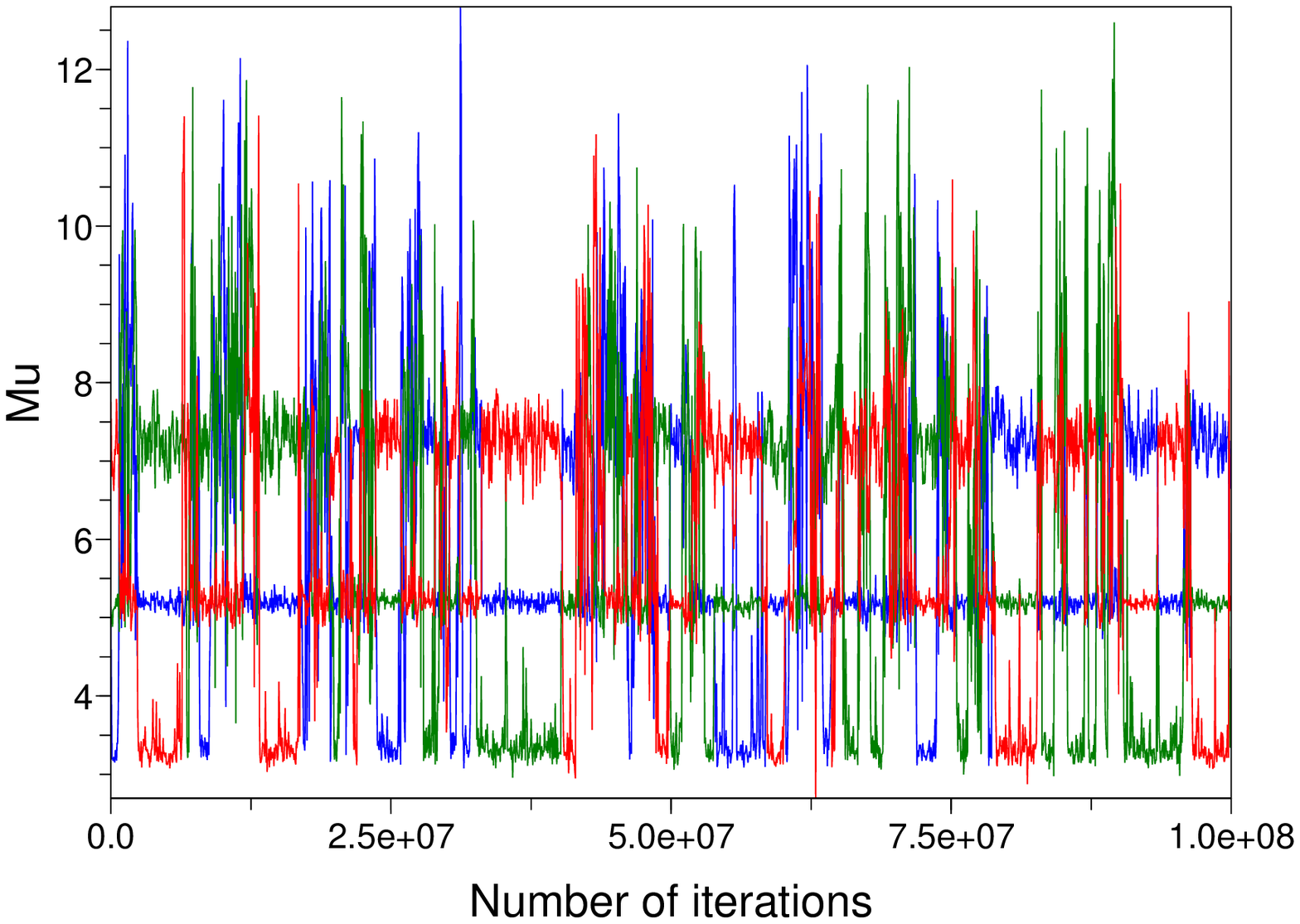}
  \includegraphics[width=7.3cm]{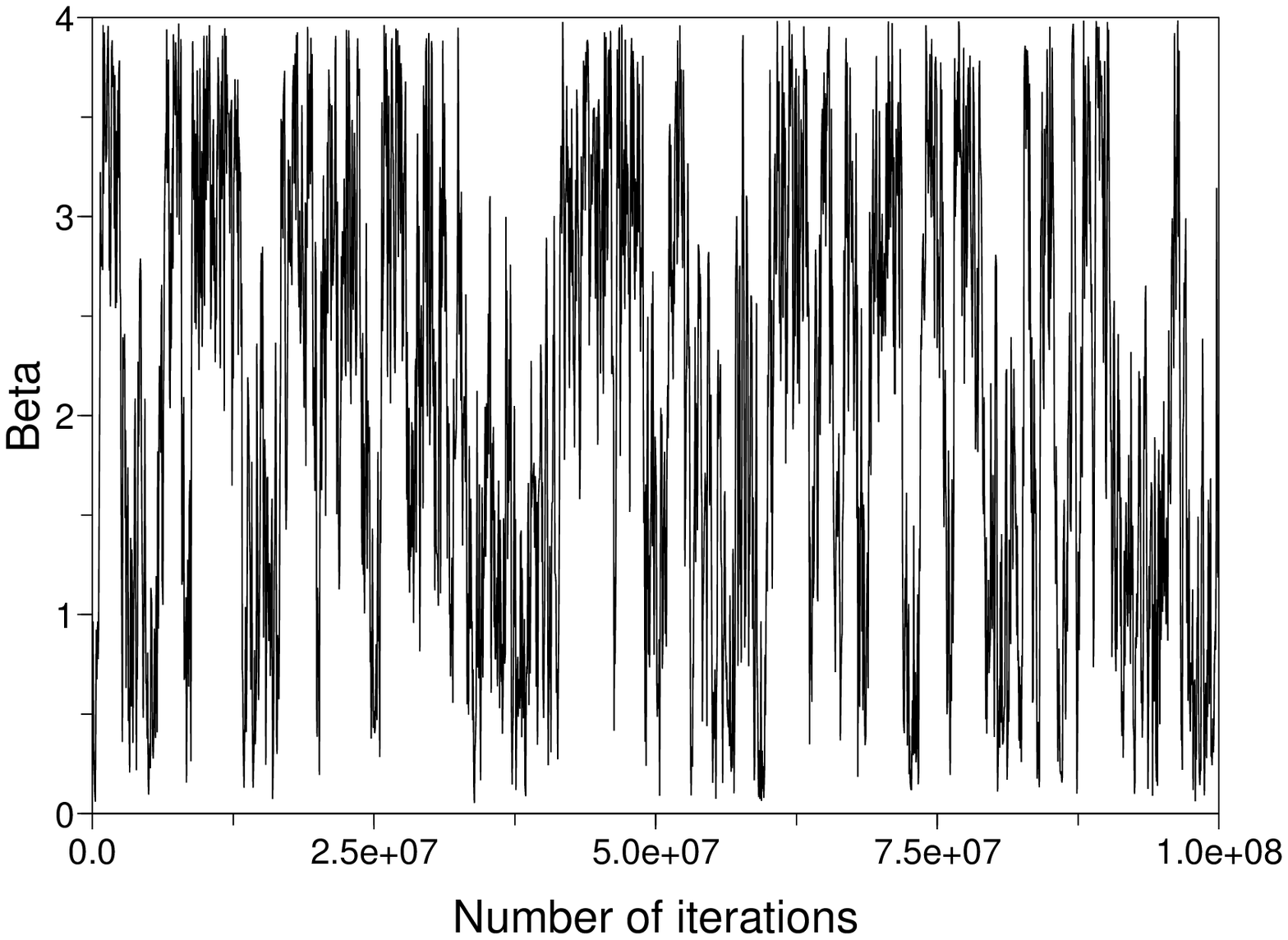}
  \caption{\label{fig:Fish3_traj} 
    Trajectories over the $10^8$ 
    first iterations of the ABF algorithm for the choice $\xi(\theta) = \beta$, 
    for $(\mu_1,\mu_2,\mu_3)$ (left) and for the $\beta$ variable (right). 
  }
\end{figure}

First, we plot on
Figure~\ref{fig:Fish3_traj} the trajectory of $(\mu_1,\mu_2,\mu_3)$
and $\beta$ for $T=10^8$ iterations. With the ABF
algorithm, the values visited by $\beta$ cover the whole interval
$[z_{\min},z_{\max}]$, and the applied bias enables a frequent
switching of the modes (observed here on the parameters
$(\mu_1,\mu_2,\mu_3)$). The trajectories for $(\mu_1,\mu_2,\mu_3)$
should be compared with the ones given on Figure~\ref{fig:no_bias},
where no adaptive biasing force is applied (note that the $x$-axis scale is not the
same on both plots). 

\begin{figure}[htbp]
  \center
  \includegraphics[width=7cm]{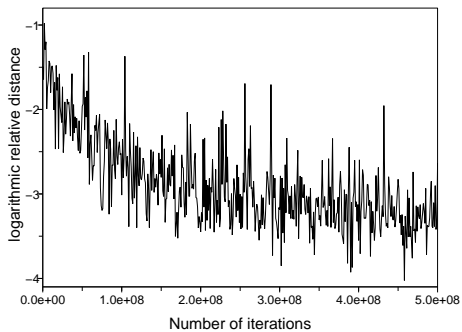}
    \caption{\label{fig:fish_cv_bias} 
      Convergence of the logarithmic relative distance
      $\log(\varepsilon_t)/\log(10)$ (see~\eqref{eq:error_bias}), 
      as a function of the number of iterations.}
\end{figure}

Second, we monitor the convergence of the biasing potential. 
To this end, we run a simulation for a total number of iterations $T = 10^9$,
and store the biasing potential every $N_{\rm cvg} = 10^6$ iterations.
The distance between the current bias
and the bias at iteration $t-N_{\rm cvg}$ is measured by
\begin{equation}
  \label{eq:error_bias}
  \delta_t = \sqrt{ \inf_{c \in \R} \sum_{i=0}^{N_z-1} \left( A_{t,i}
      - A_{t-N_{\rm cvg},i} - c\right)^2 },
\end{equation}
where $A_{t,i}$ denotes the value of the bias in bin $i$,
\emph{i.e.}  $A_t(z)=A_{t,i}$ if $z\in(z_{i},z_{i+1})$. 
Since the potential is defined only up to an additive constant, we consider the optimal
shift constant $c$ which minimizes the mean-squared distance between the two profiles.
An elementary computation shows that this constant is equal to the difference between the
averages of $A_{t}$ and $A_{t-N_{\rm cvg}}$.
We finally renormalize this distance as
\[
\varepsilon_t = \frac{\delta_t}{\sqrt{ \sum_{i=0}^{N_z-1}A_{t,i}^2}}.
\]
The relative distance $\varepsilon_t$ as a function of the iteration index $t$ is
plotted in Figure~\ref{fig:fish_cv_bias}. Correct approximations of the bias
are obtained after a few multiples of $N_{\rm cvg}$ iterations
(the relative error being already lower than 0.1 at the first convergence check).
We emphasize again that we did not optimize the proposal moves in order to reach
the fastest convergence of the bias. It is very likely that  better convergence
results could be obtained by carefully tuning the parameters of the 
proposal function, or resorting to proposals of a different type.

\begin{figure}[htbp]
  \center
  \includegraphics[width=4cm,angle=270]{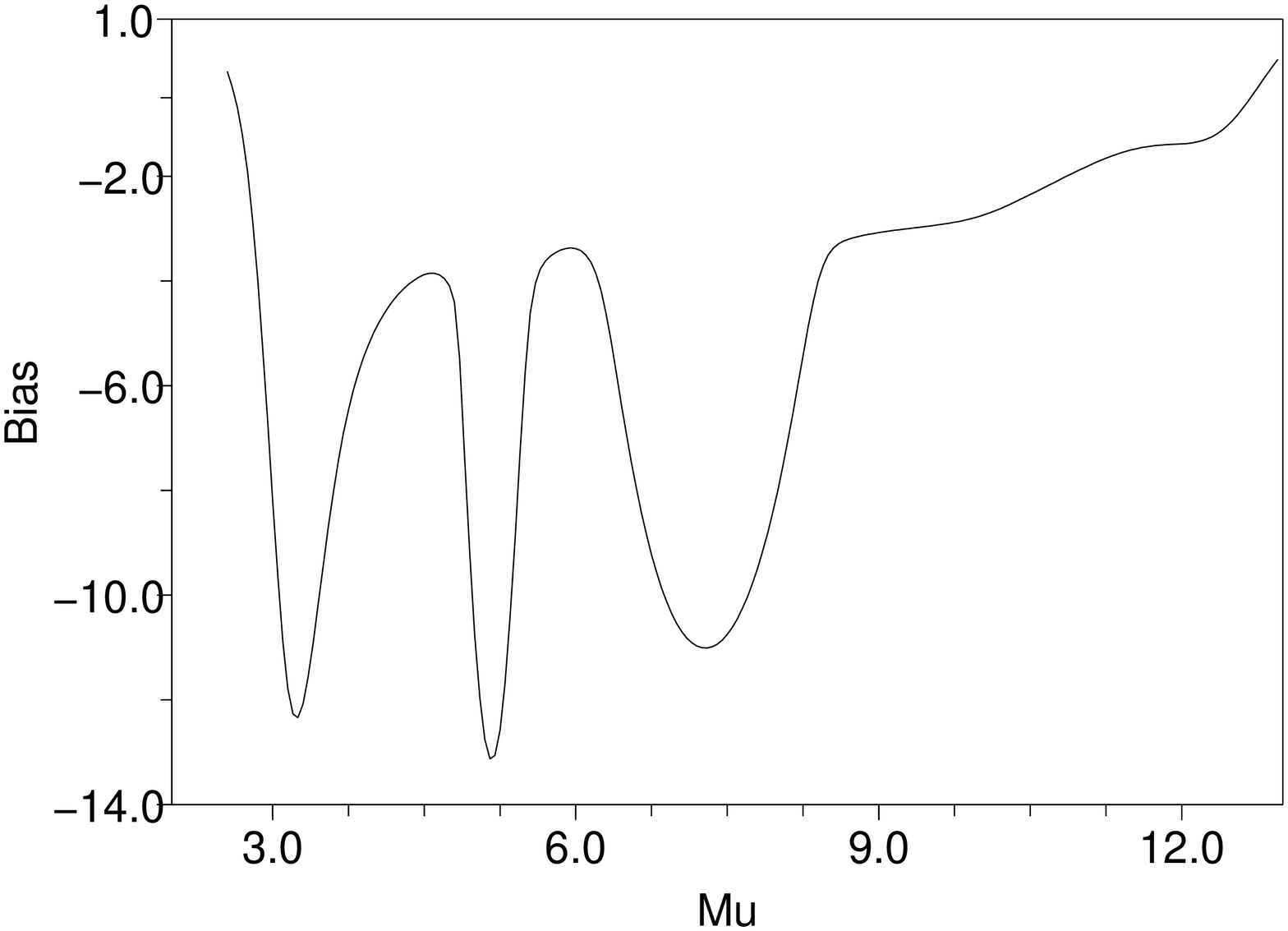}  
  \includegraphics[width=4cm,angle=270]{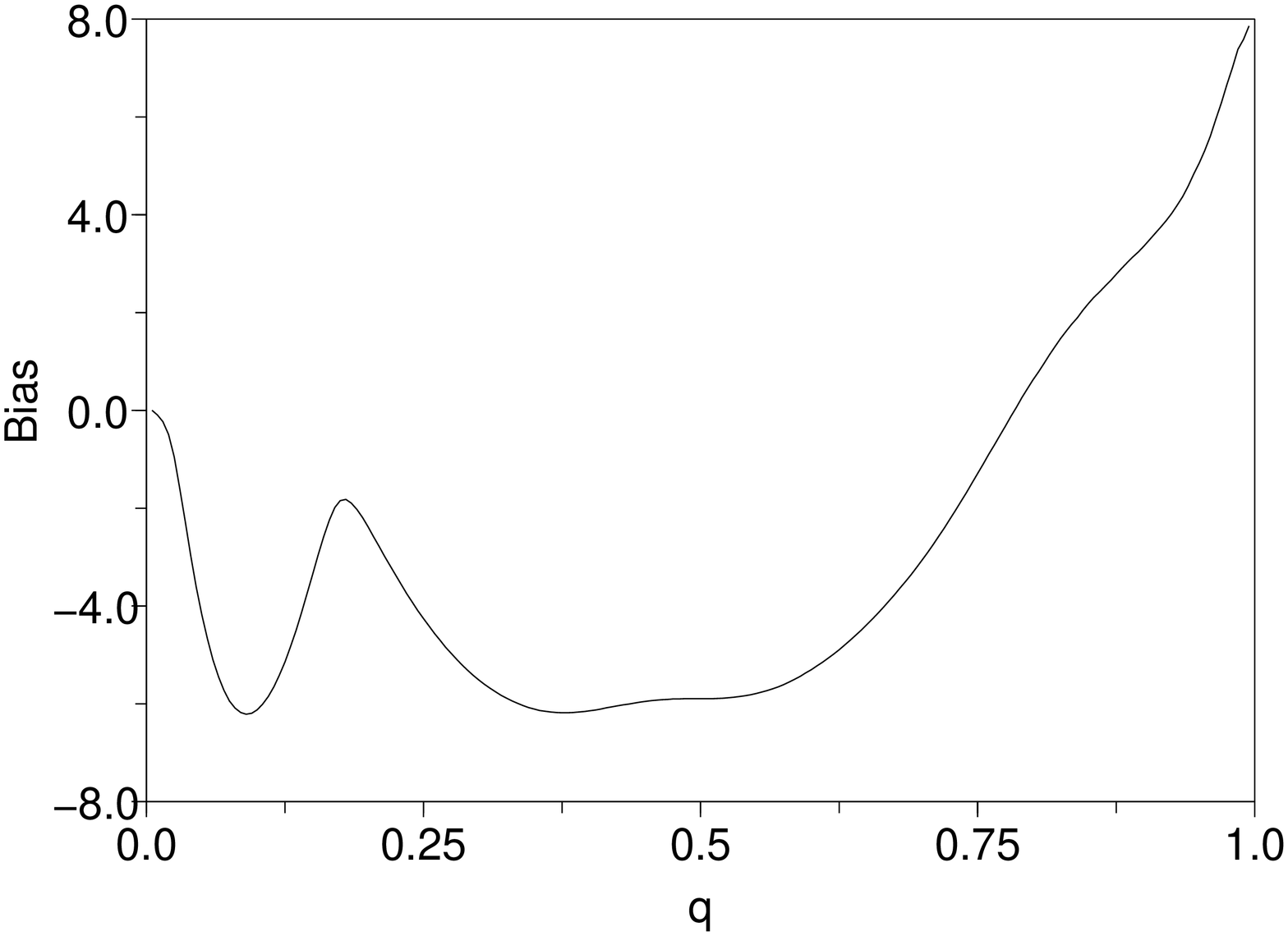}\\
  \includegraphics[width=4cm,angle=270]{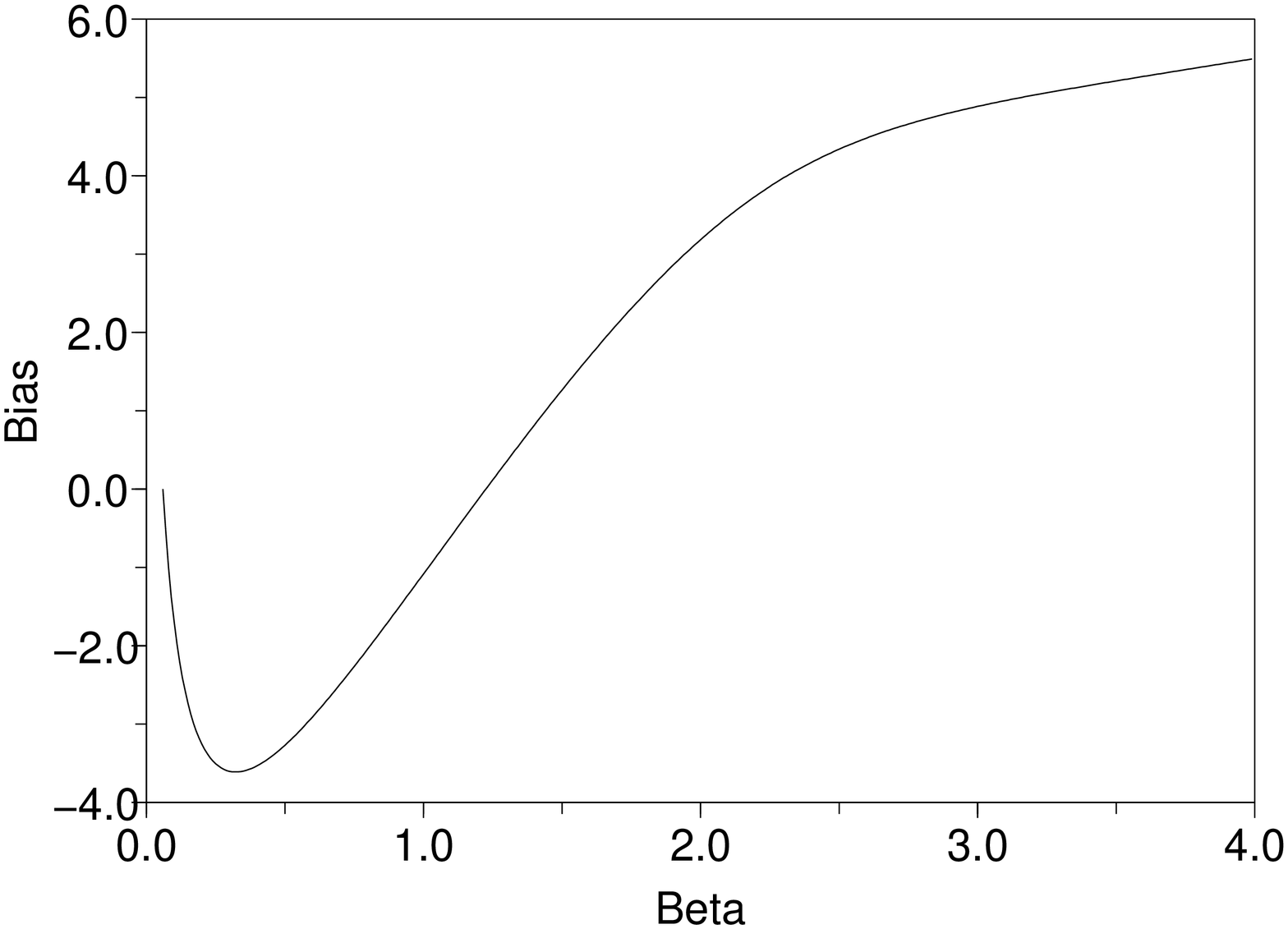}
  \includegraphics[width=4cm,angle=270]{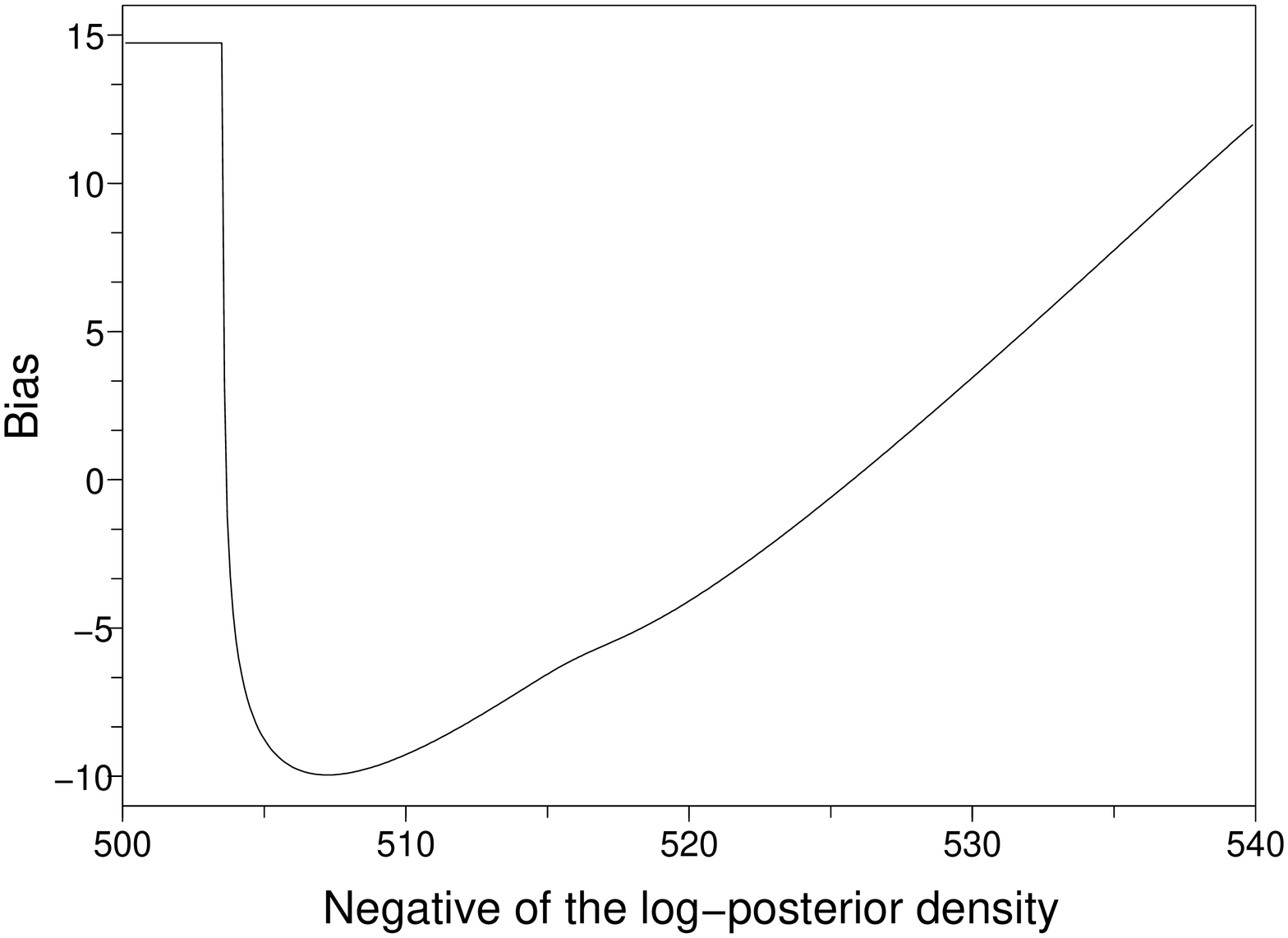}
  \caption{\label{fig:fish_bias} 
    The fishery data. Free energies obtained for the reaction coordinates: 
    $\xi(\theta)=\mu_1$ (top left), $\xi(\theta)=q_1$ (top right), 
    $\xi(\theta)=\beta$ (bottom left) and $\xi(\theta)=-\log\{p(\theta)p(y|\theta)\}$ 
    (bottom right).}
\end{figure}

On Figure~\ref{fig:fish_bias}, we plot the free energies associated to
the four reaction coordinates mentioned above, as estimated by
adaptive algorithms. 
We recall that, for $\xi(\theta) = \beta$, $z_{\rm min} = 0.05$,
$z_{\rm max} = 4.0$ and $\Delta z = 0.01$.
For $\xi(\theta)=-\log\{p(\theta)p(y|\theta)\}$,
we used an ABP algorithm, with $z_{\rm min} = 500$, $z_{\rm max} =
540$ and $\Delta z = 0.1$.  For $\xi(\theta) = q_1$ and $\xi(\theta) =
\mu_1$, we used ABF, with respectively $z_{\rm min} = 0$, $z_{\rm max}
= 1$, $\Delta z = 0.005$ and $z_{\rm min} = \min y_i = 2.5$, $z_{\rm
  max} = \max y_i = 13$, $\Delta z = 0.05$. Recall that the
so-obtained bias is minus the marginal posterior log-density of~$\xi$. This
is why the three important modes in the $\mu$ parameter can be read
from the corresponding bias in Figure~\ref{fig:fish_bias}.  Note
also that there is a lower bound on the admissible values of minus the
log-posterior density, hence the plateau value of the corresponding
bias for low values of the reaction coordinate corresponding to unexplored regions.

\subsubsection{Efficiency of the biasing procedure}
\label{sec:eff_samp_num}

\begin{figure}[htbp]
  \center
  \includegraphics[width=7.3cm]{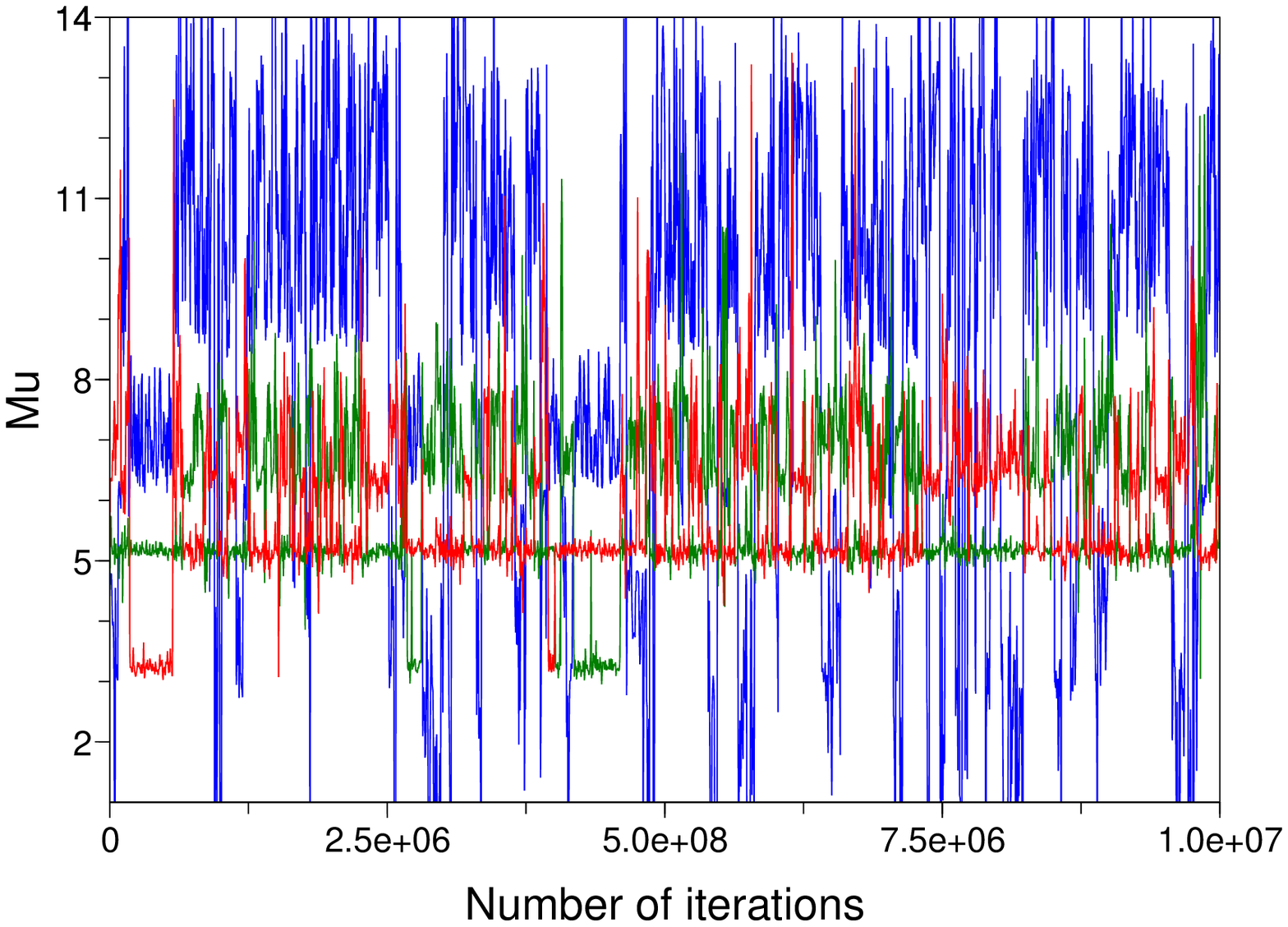}  
  \includegraphics[width=7.3cm]{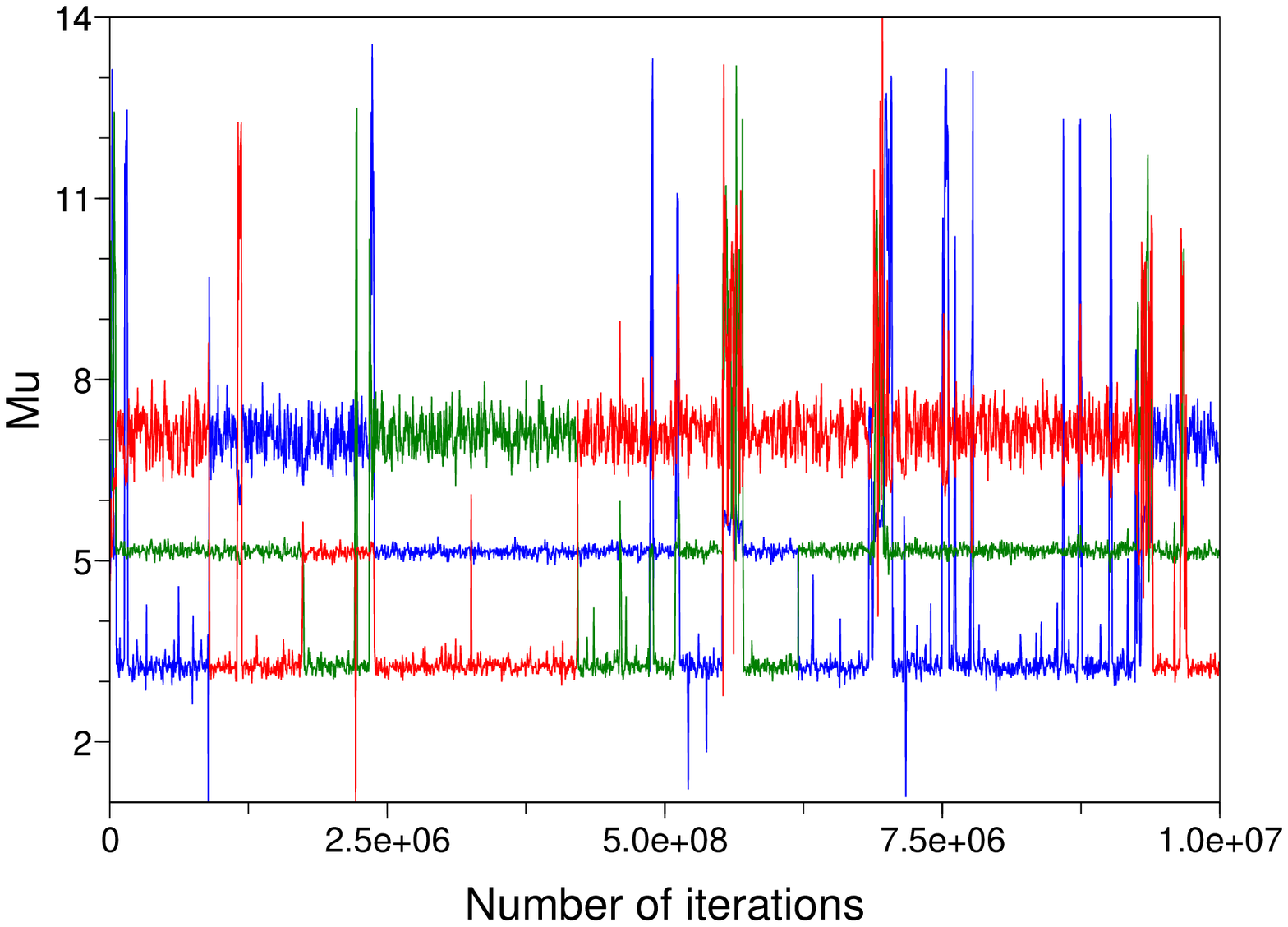} \\
  \includegraphics[width=7.3cm]{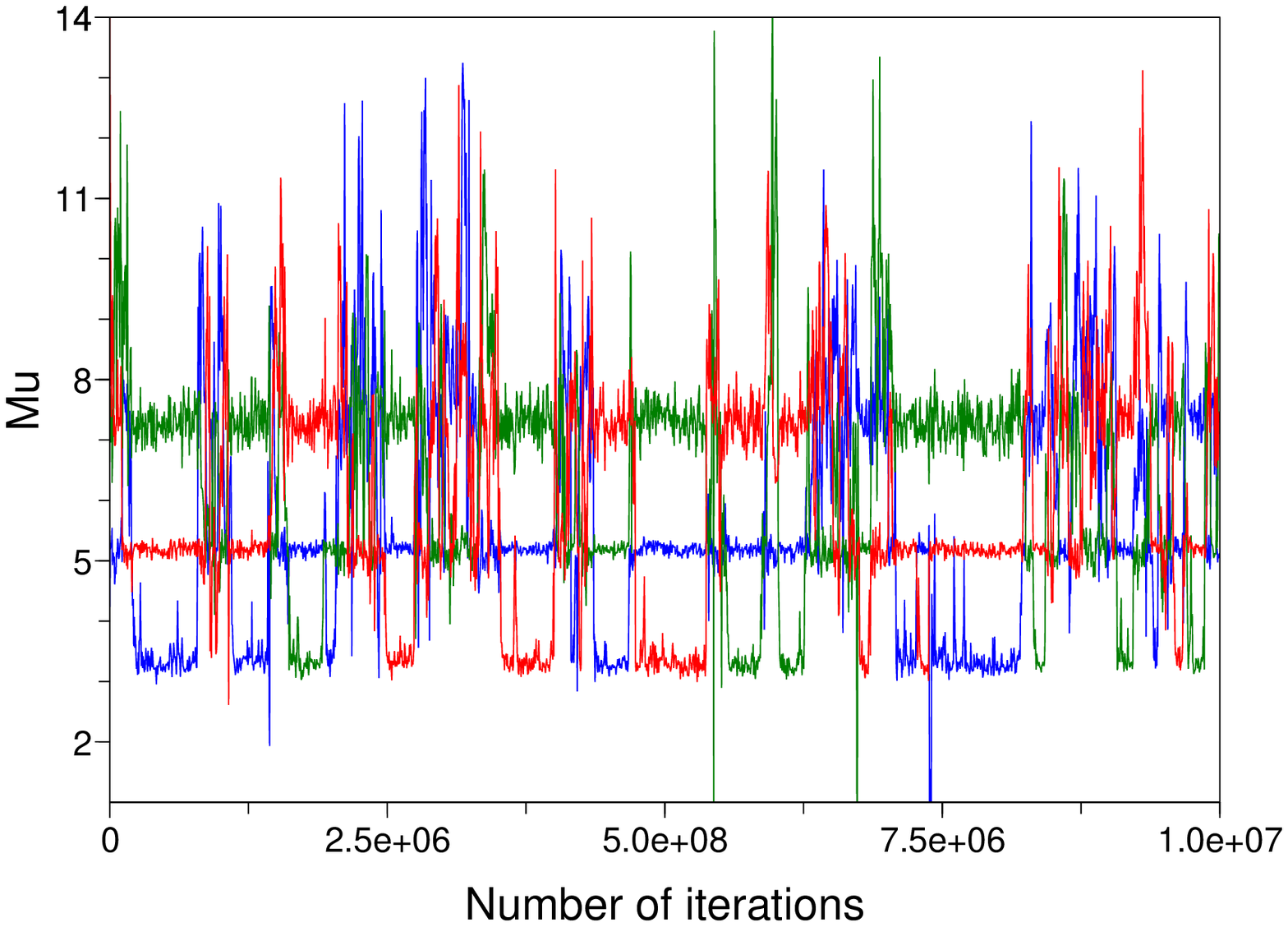} 
  \includegraphics[width=7.3cm]{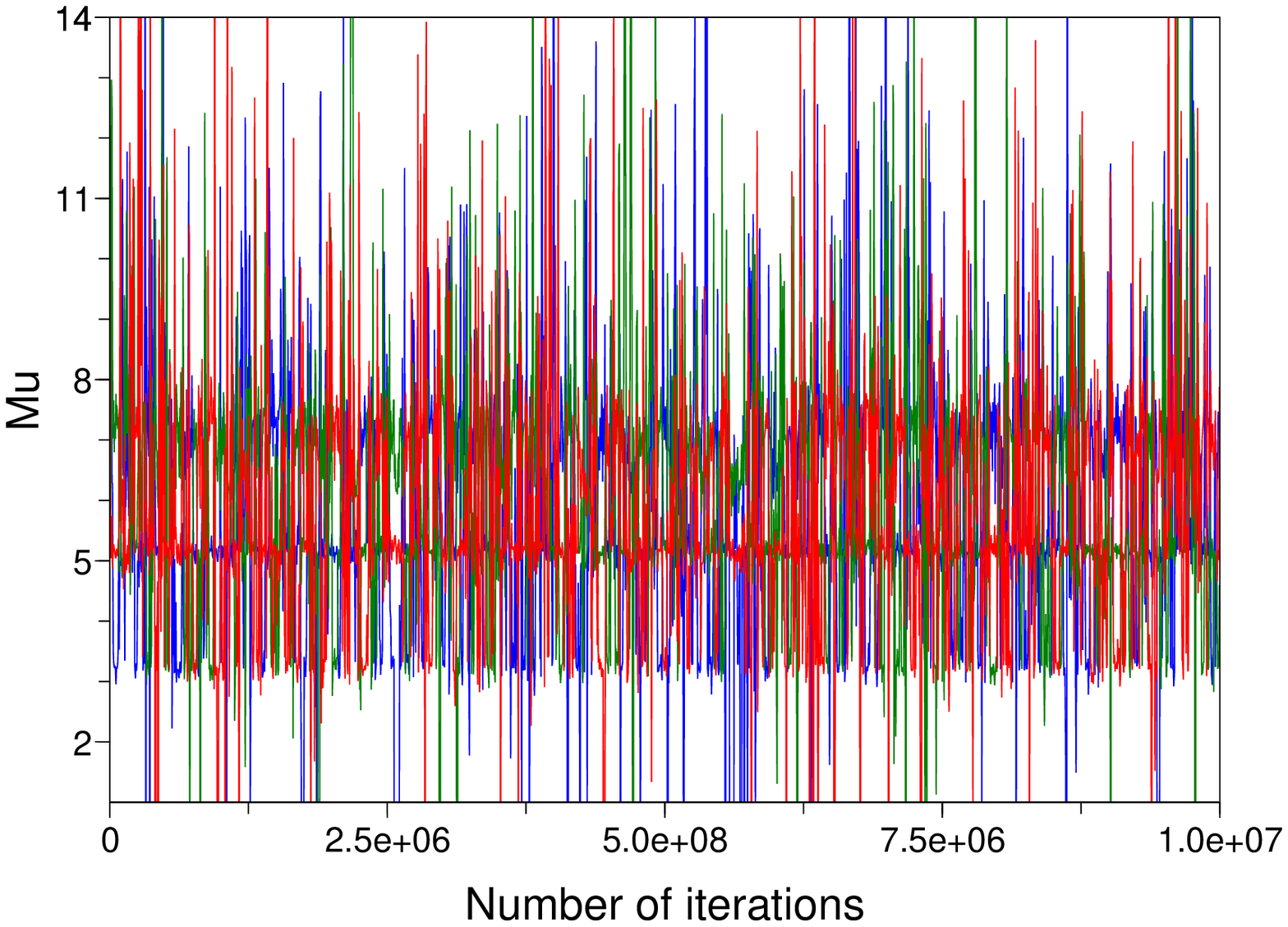}  
  \caption{\label{fig:fish_bias_traj} 
    The fishery data. Trajectories of $(\mu_1,\mu_2,\mu_3)$ 
    for the biased dynamics for different reaction
    coordinates.
    Top left: $\xi(\theta)=\mu_1$. Top right: $\xi(\theta)=q_1$.
    Bottom left: $\xi(\theta)=\beta$.
    Bottom right: $\xi(\theta)=-\log\{p(\theta)p(y|\theta)\}$.
  }
\end{figure}

We now discuss the results of the MCMC algorithm targeted at the
biased posterior distribution, using the free energies computed above. 
In
Figure~\ref{fig:fish_bias_traj}, we observe that all the biased dynamics 
are much more successful in terms of mode switchings than the unbiased dynamics
(see Figure~\ref{fig:no_bias}).
More precisely, the dynamics biased by the free energy associated with
$\xi(\theta)=-\log\{p(\theta)p(y|\theta)\}$ is the most successful in terms
of switchings, but the dynamics with $\xi(\theta) = \beta$ performs correctly as
well. The dynamics  with $\xi(\theta) = q_1$ seems to be less successful.
In the case $\xi(\theta) = \mu_1$, one value of the parameters $\mu$ is forced to 
visit the whole range of values. The lowest mode (around $\mu=3$) is not 
very well visited here.

The efficiency factors are presented in
Table~\ref{tab:fish_EF_RC}.  They are rather large, which shows that
the importance sampling procedure does not yield a 
degenerate sample. The choice $\xi(\theta) = q_1$ is the best one,
but $\xi(\theta) = \beta$ and $\xi(\theta) = -\log\{p(\theta)p(y|\theta)\}$
give comparable and satisfactory results as well. The choice 
$\xi(\theta) = \mu_1$ on the other hand is a poor choice in this case.

In view of these results, it seems that $\xi(\theta) =
-\log\{p(\theta)p(y|\theta)\}$ is the best choice, with the problem
however that we had to slightly modify the bias for the lowest
values of the reaction coordinate because of the too sharp variations
of the bias in this region. (Our numerical experience is indeed that the bias
obtained from minus the log-posterior density is sometimes
difficult to use directly.) On the other hand, the procedure is
more automatic for $\xi(\theta) = q_1$ and $\beta$, the latter
reaction coordinate being a much better choice when it comes to mode
switchings.

\begin{table}
\begin{center}
\begin{tabular}{c|c|c|c|c}
Reaction coordinate & $\beta$ & $-\log\{p(\theta)p(y|\theta)\}$ & $q_1$ & $\mu_1$ \\
\hline
EF (numerical)   & 0.17  & 0.16  & 0.48 & 0.04 \\
EF (theoretical) & 0.179 & 0.178 & 0.454 & 0.079 \\   
\end{tabular}
\end{center}
\caption{Efficiency factor for various choices of reaction coordinates, in the case
$K=3$.}
\label{tab:fish_EF_RC}
\end{table}

\begin{figure}[htbp]
  \center
  \includegraphics[width=7.3cm]{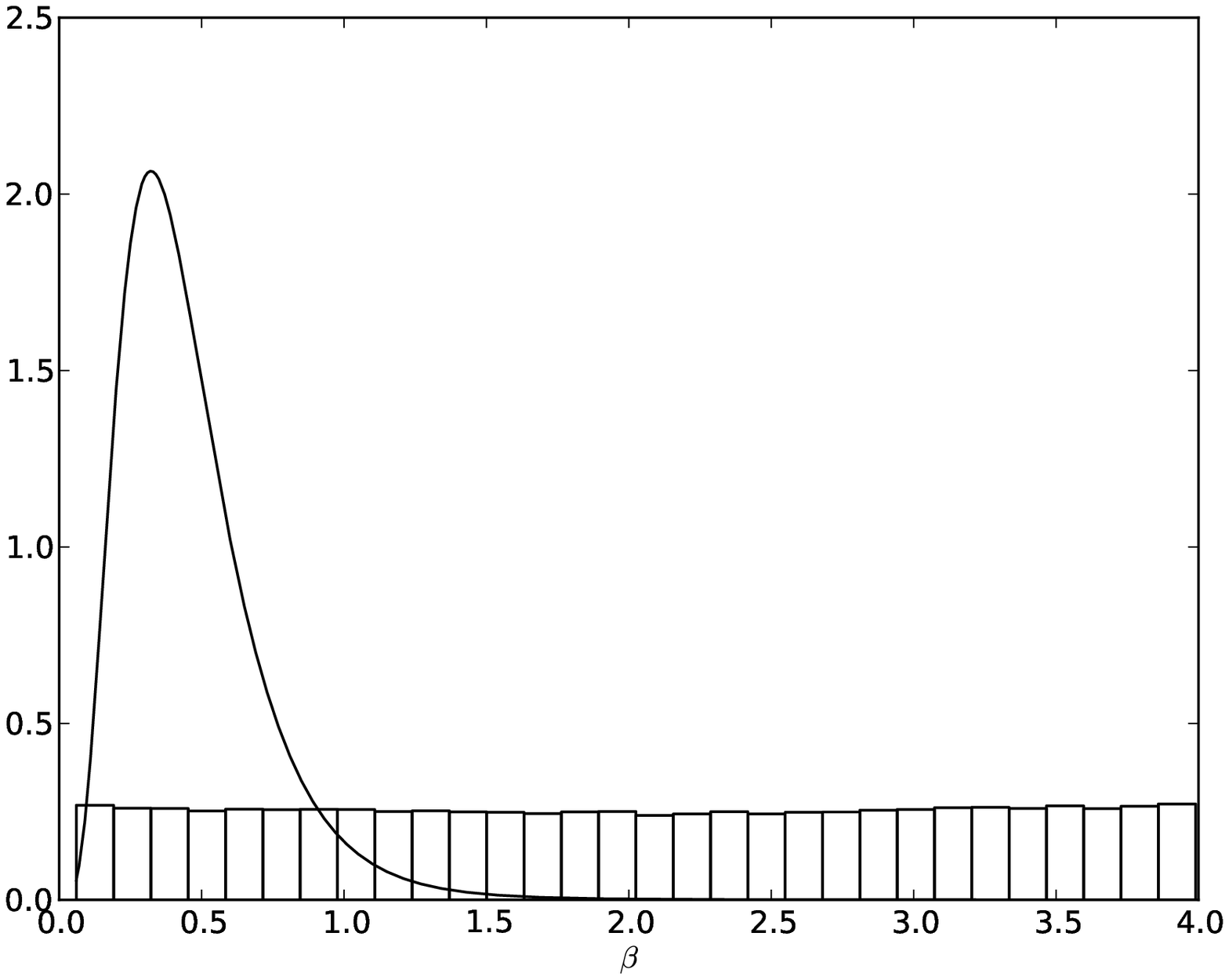}
  \includegraphics[width=7.3cm]{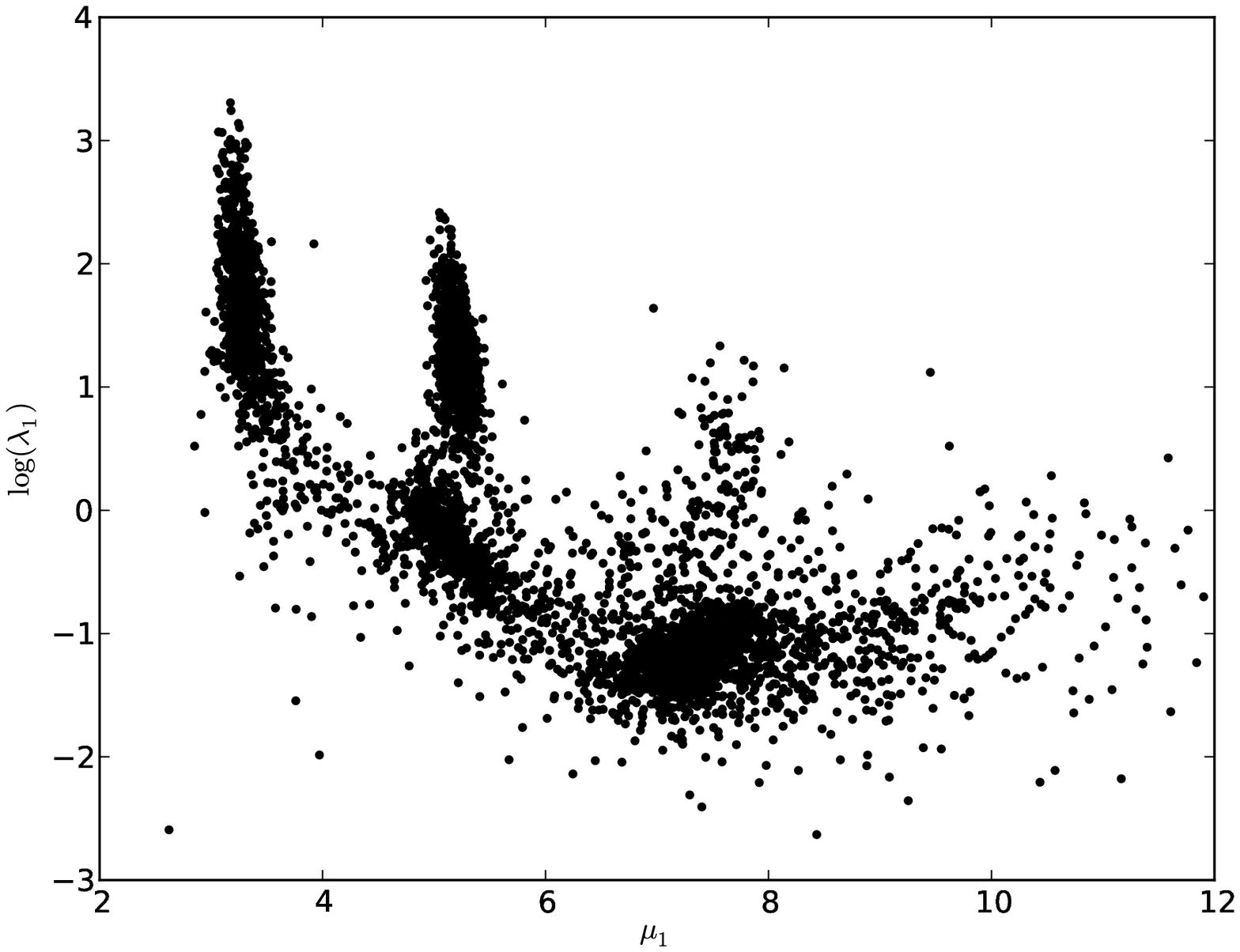}\\
  \includegraphics[width=7.3cm]{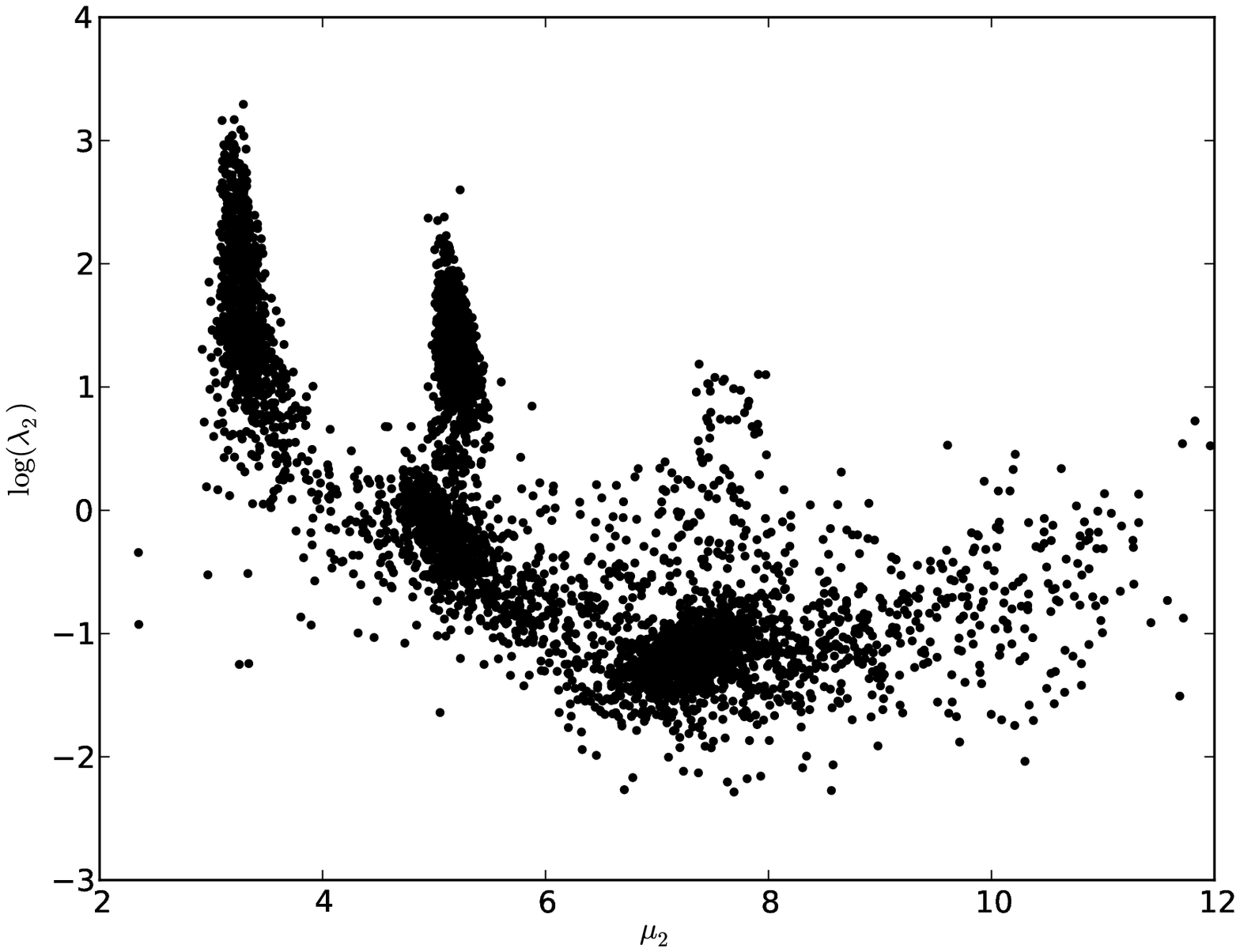}
  \includegraphics[width=7.3cm]{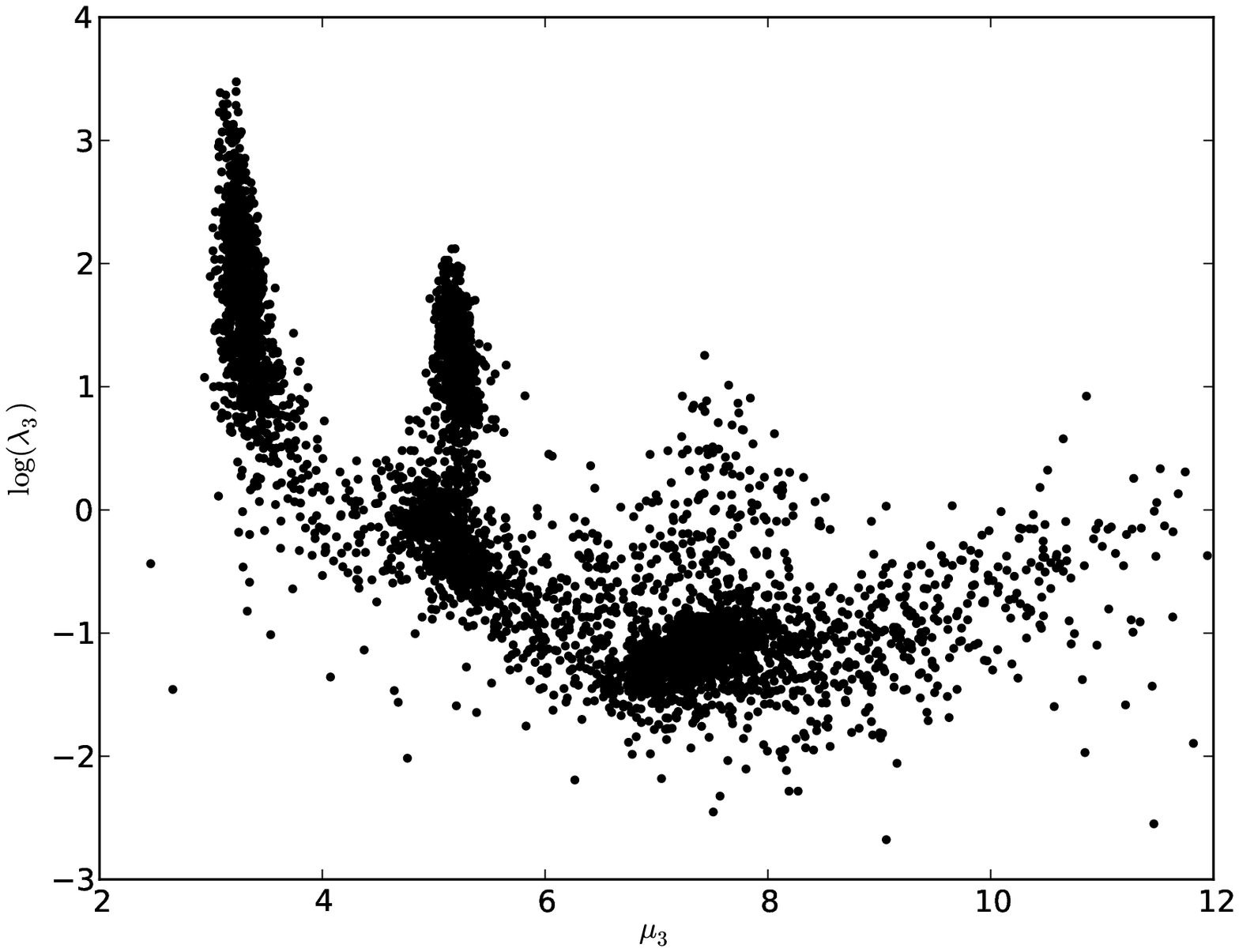}
    \caption{\label{fig:fish_muv} Top left: Histogram of simulated $\beta$'s 
      and estimated marginal posterior density of $\beta$. Remaining pictures: Scatter 
      plots of simulated values for $(\mu_i,\log\lambda_i)$, for $i=1,2,3$ 
      when $\beta$ is used as a reaction coordinate.}
\end{figure}

We now focus on $\xi(\theta)=\beta$. As explained in the introduction,
a good sampler should visit all the possible labellings of the
parameter space.  This implies in particular that the marginal
posterior distributions of the simulated component parameters should
be nearly identical. This is clearly the case here, see the
scatter plots of the 1-in-$10^4$ sub-sample of the simulated pairs
$(\mu_i,\log \lambda_i)$, $i=1,\ldots,3$ in Figure \ref{fig:fish_muv}.
The top left picture in Figure~\ref{fig:fish_muv} also demonstrates
that the biased dynamics indeed samples uniformly the values of
$\beta$ over the chosen interval $[z_{\rm min}, z_{\rm max}]$.

\begin{figure}[htbp]
  \centering
  \includegraphics[width=16cm,trim=3cm 0 0 0]{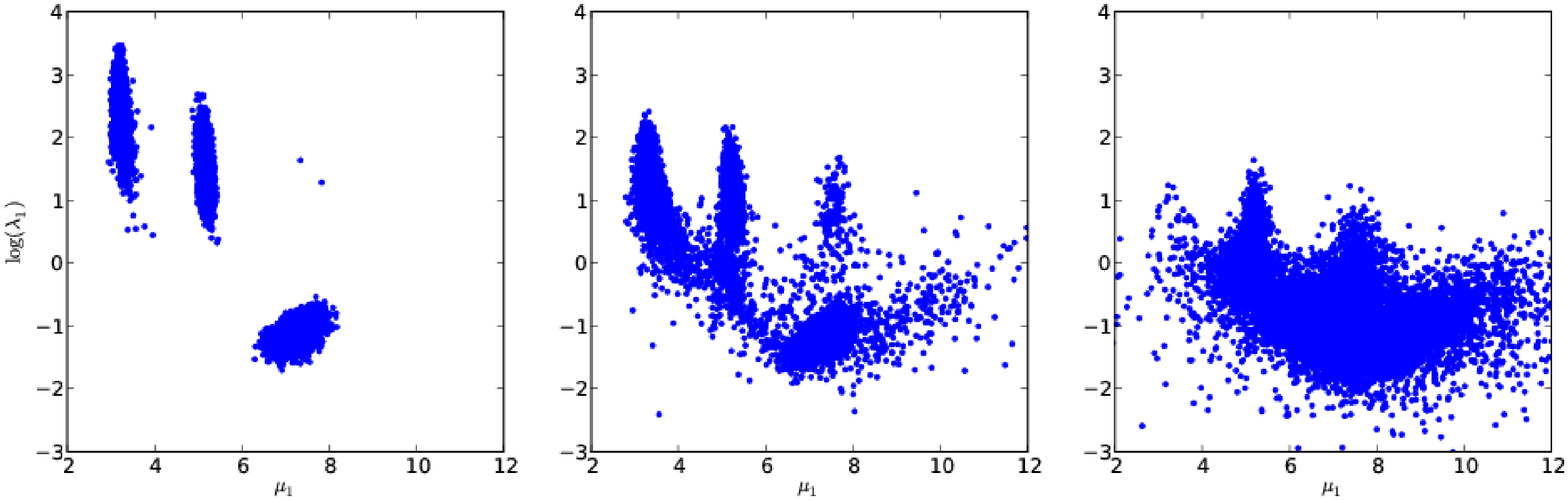}
  \caption{Simulated pairs $(\mu_1,\log\lambda_1)$ conditional on,
    from left to right, $\beta\in[0,0.5]$, $\beta\in[1.5,2]$ and
    $\beta\in[3.5,4]$, see Section~\ref{sec:appl-fish-data} for more details.}
  \label{fig:mulambda_condbeta}
\end{figure}

Finally, Figure~\ref{fig:mulambda_condbeta} illustrates why the
reaction coordinate $\xi(\theta)=\beta$ allows for escaping from local
modes; see the discussion in Section
\ref{sec:react-coord-mixt}. 
Each plot represents a sub-sample of the simulated pairs
$(\mu_{k,t},\log\lambda_{k,t})$ (where the subscript $t$ denotes
the iteration index while the subscript $k$ labels the components), 
restricted to $\beta_t$ being in intervals,
from left to right, $[0,0.5]$, $[1.5,2]$ and $[3.5,4]$. Since the bias
function depends only on $\beta$, these plots are rough approximations
of the posterior distribution conditional on $\beta=0.25$ (its posterior expectation),
$\beta=1.75$ and $\beta=3$. 
In the leftmost plot
of Figure \ref{fig:mulambda_condbeta}, $\beta$ is fixed to its posterior expectation
and the three modes are well separated. As
$\beta$ is forced to take artificially large values (in the sense that
the posterior probability density of such values is very
small), the three modes get closer and eventually merge. 

\subsubsection{Larger values of $K$ and model choice}
\label{sec:appl-fish-data}

\begin{figure}[htbp]
  \center
  \includegraphics[width=7.3cm]{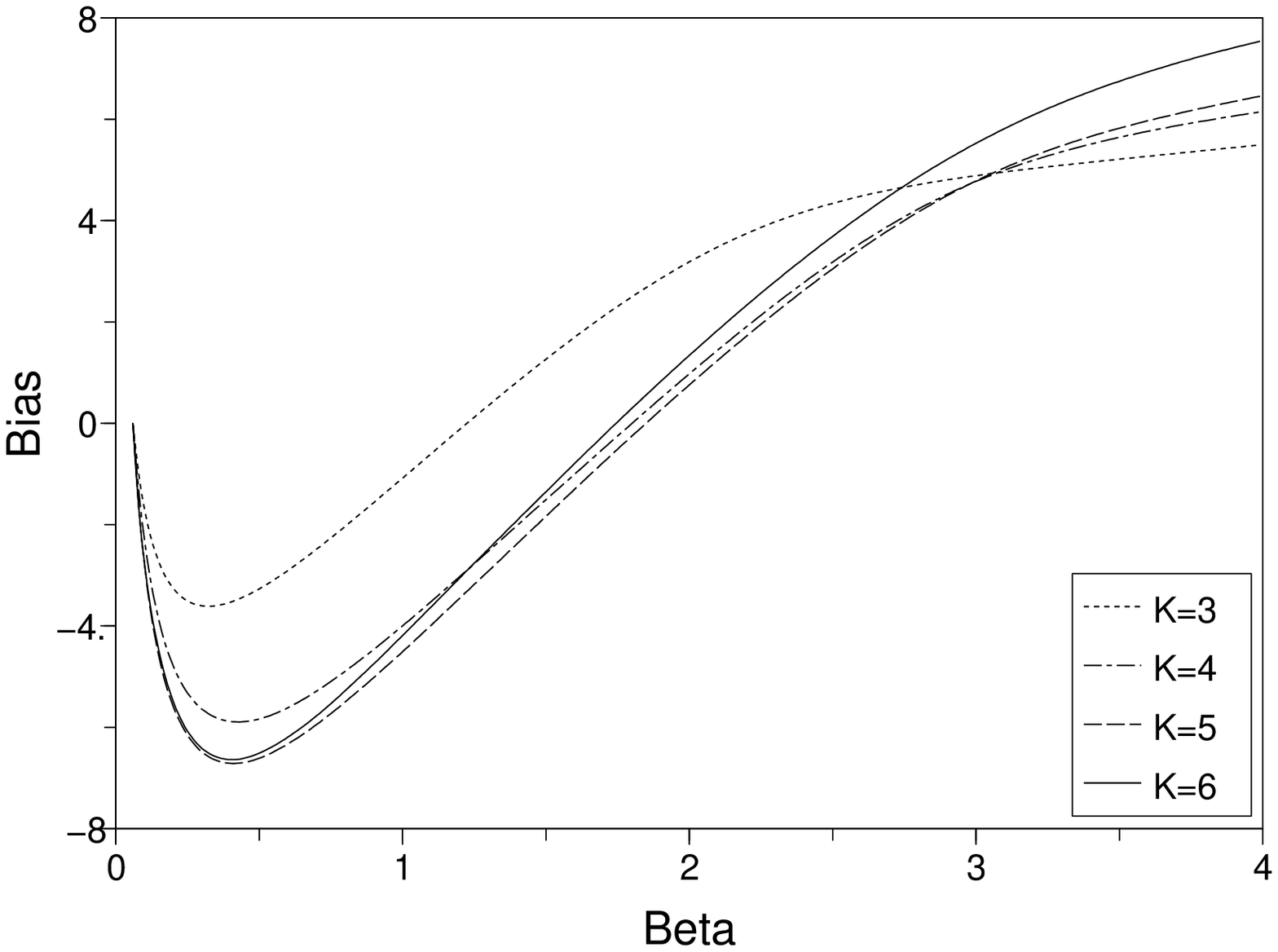}
  \caption{\label{fig:fish_bias_K3to6} Estimated bias (free energy), for
      $K=3,\ldots,6$, and $\xi(\theta)=\beta$.}
\end{figure}

We apply our approach to other values of $K$, namely $K=4$ to 6, 
in the case $\xi(\theta)=\beta$.
Table~\ref{tab:fish_EF} reports the efficiency factor as a function of 
$K$. These factors remain quite satisfactory, 
which  is related to the fact that the amplitude of the
free energy (difference between the maximum and the minimum values)
associated to this reaction coordinate is not too large
over the chosen interval $[z_{\rm min}, z_{\rm max}]$, and does not change
dramatically, see the profiles
obtained for different values of $K$ in Figure~\ref{fig:fish_bias_K3to6}. 

\begin{table}
\begin{center}
\begin{tabular}{c|c|c|c|c}
$K$ & 3 & 4 & 5 & 6 \\
\hline
EF (numerical) & 0.17 & 0.18 & 0.17 & 0.16 \\
EF (theoretical) & 0.179 & 0.195 & 0.180 & 0.171 \\
\end{tabular}
\end{center}
\caption{Efficiency factors, for various values of the number of 
components considered in the mixture, with the choice $\xi(\theta) = \beta$.}
\label{tab:fish_EF}
\end{table}

\begin{figure}
  \centering
  \includegraphics[scale=0.35,trim=4cm 0 0 0]{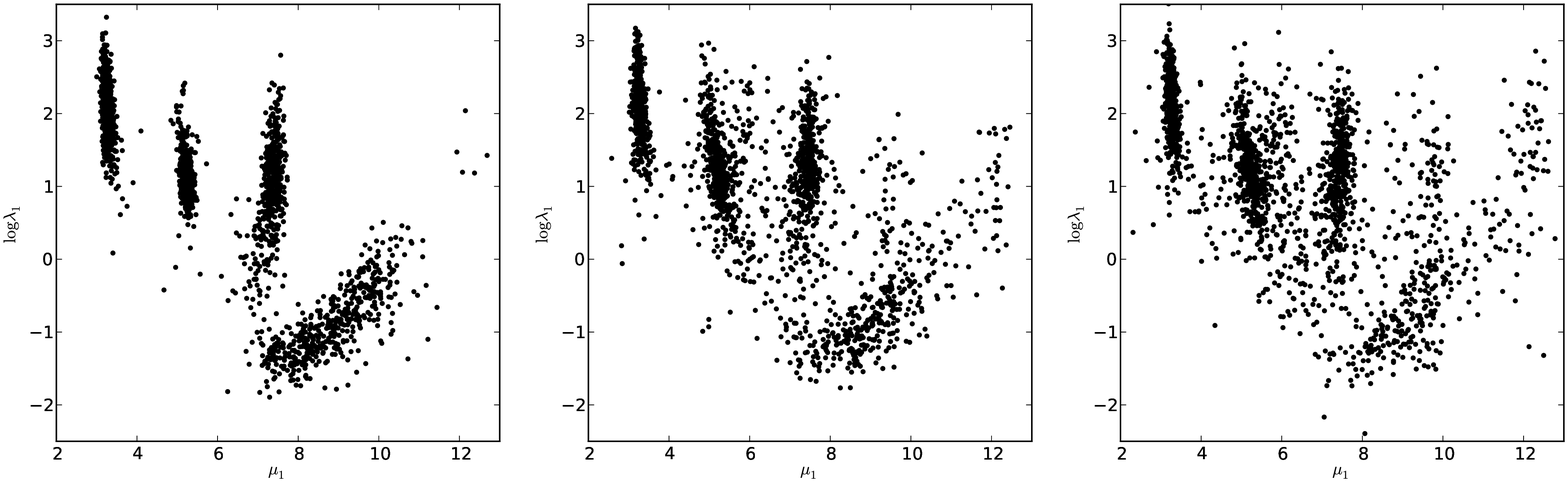}
  \caption{Marginal posterior distribution of $(\mu_1,\log\lambda_1)$,
    from left to right, for $K=4$, 5 and 6, as represented by 2000
    points resampled from the MCMC output.}
  \label{fig:Fish456}
\end{figure}

Figure \ref{fig:Fish456}
represents the marginal posterior distribution of
$(\mu_1,\log\lambda_1)$, for $K=4,5,6$. These plots are obtained
by resampling 2000 points from the output of the MCMC targeting the
biased posterior, with probability
proportional to the importance sampling weight defined in
\eqref{eq:wis}. In each case, we checked that the MCMC output is symmetric with
respect to label permutations. 

Table \ref{tab:fish} reports, for $K=3,\ldots,6$, the estimated
log-Bayes factor for choosing a mixture model with $K$ components
against a mixture model with $K-1$ components, which equals
$\log Z_K/Z_{K-1}$, assuming equal prior probability for different
values of $K$. The reported error levels in Table~\ref{tab:fish}
correspond to $90\%$ confidence intervals, which are deduced from
repeated independent runs of $T=10^7$ iterations from the same MCMC algorithm
targeting the biased posterior. The estimation error is quite small,
despite being based on importance sampling steps in high 
dimensional spaces.

\begin{table}
\begin{center}
\begin{tabular}{c|c|c|c|c}
$K$ & $\log Z_K/Z_{K-1}$ & error \\ 
\hline 
3 & 7.1 & $\pm$  0.2\\ 
4 &  4.2 &  $\pm$  0.1 \\
5 & 1.5 &  $\pm$ 0.1 \\
6 & 0.9 &  $\pm$ 0.1 \\
\end{tabular}
\end{center}
\caption{Log Bayes factors for comparing model with $K$ components
  against $K-1$ components, for $K=3,...,6$; estimation error as
  evaluated from independent MCMC runs.}
\label{tab:fish}
\end{table}


\subsection{A second example: the Hidalgo stamps data}
\label{sec:hidalgo}

Another well-known benchmark for mixtures is the Hidalgo stamps
dataset, first studied by~\cite{IS88} (see also \emph{e.g.}~\cite{BMY97}),
which consists of the thickness (in mm) of $n=485$ stamps from a given
Mexican stamp issue; see Figure~\ref{fig:no_bias} for a
histogram. (For convenience we multiplied the observations by 100.) We
focus our presentation on the challenging case $K=3$. 
For other values of $K$ between 4 and 7 our approach performs 
better than for $K = 3$. For the sake of space the
corresponding results are not reported. 

\begin{figure}[htbp]
  \center
  \includegraphics[width=7.3cm]{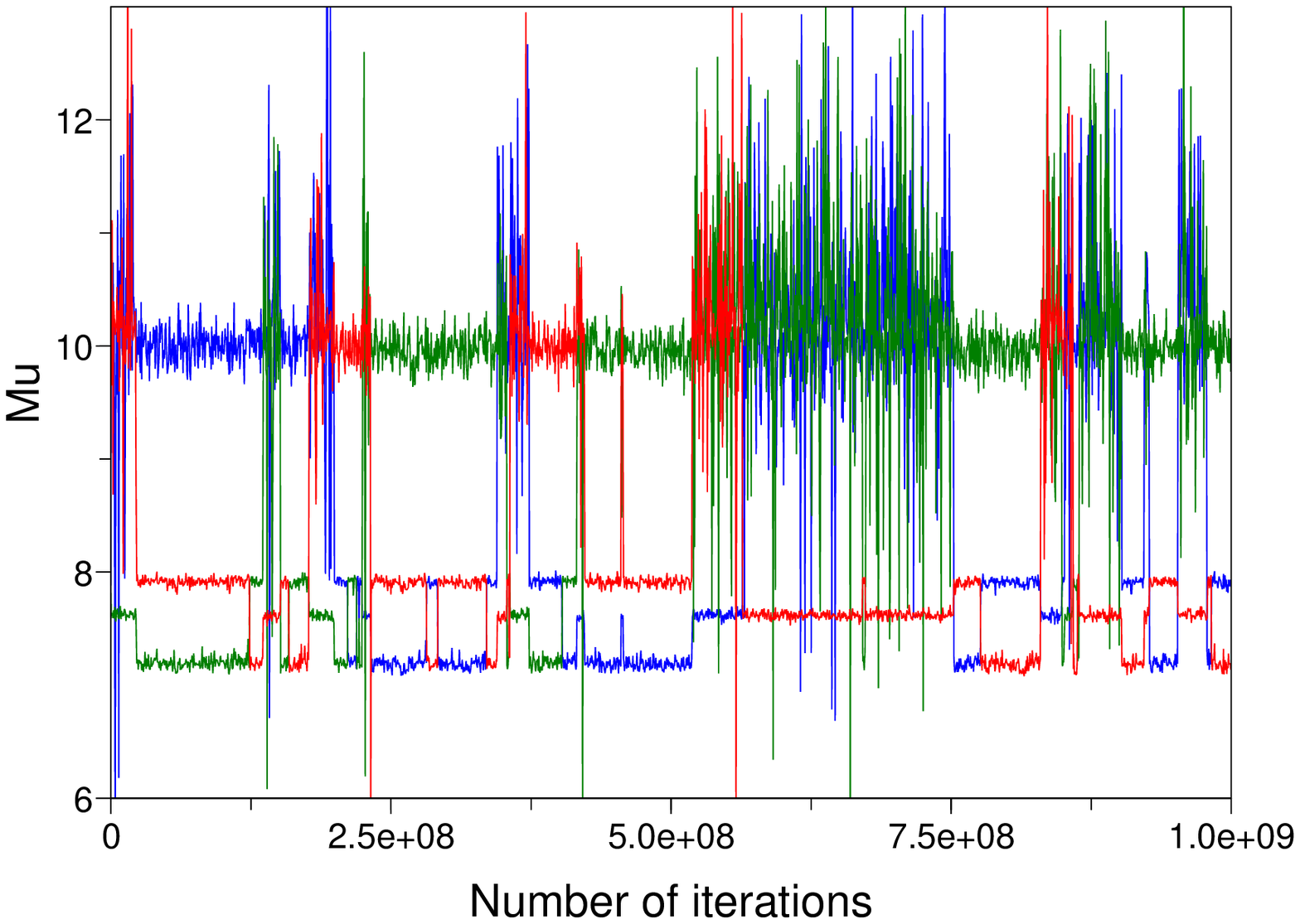}
  \includegraphics[width=7.3cm]{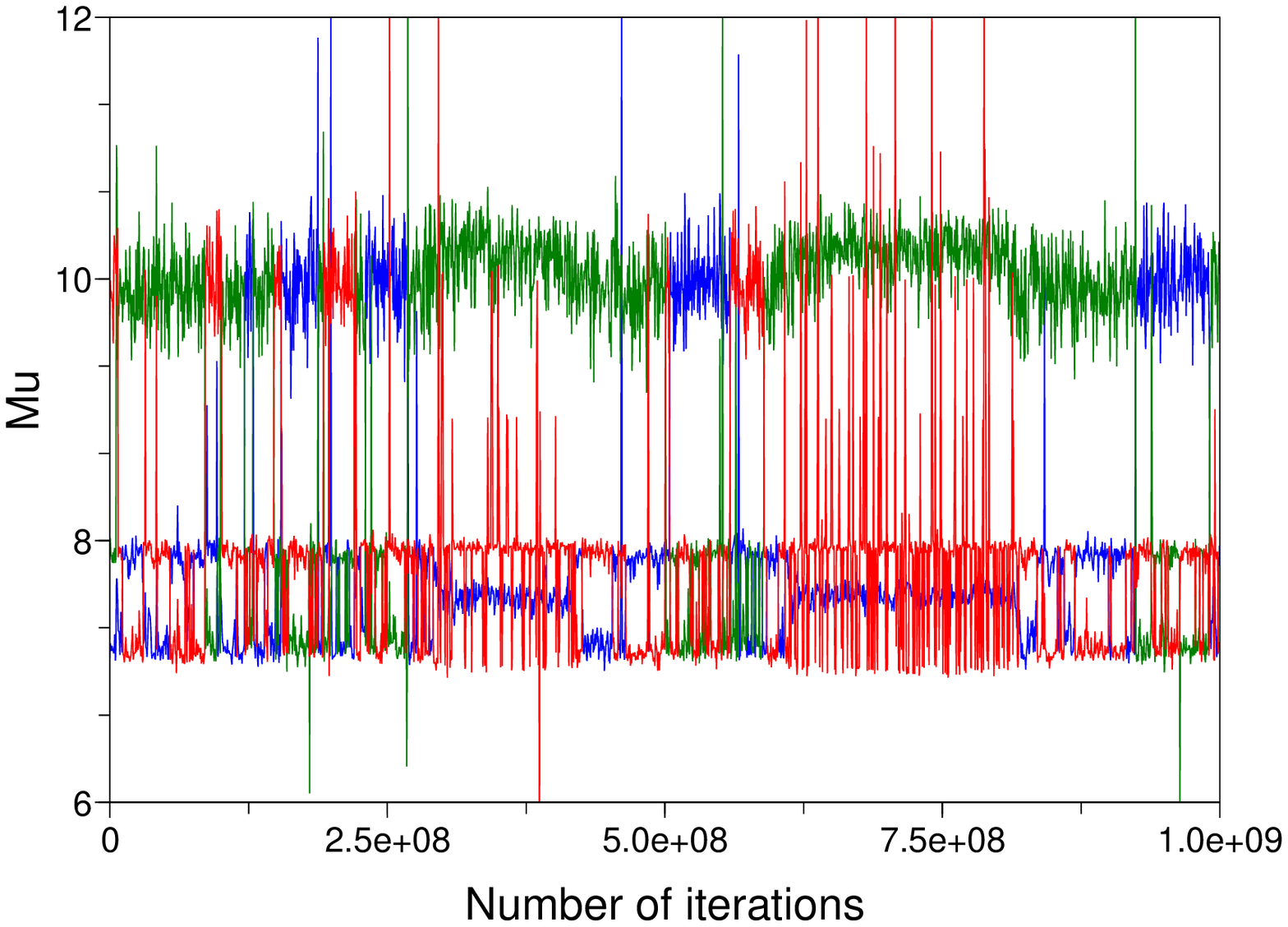}
  \caption{\label{fig:Hidalgo_beta} Sampled trajectories for
    $(\mu_1,\mu_2,\mu_3)$ during the adaptive biasing procedure.
    Left: ABF trajectory when the reaction coordinate is
    $\beta$. Right: ABP trajectory when the reaction coordinate is 
    minus the log-posterior density.}
\end{figure}

This example is more challenging than the previous one,
presumably because the number of observations is larger, which makes
the likelihood more peaked.  A clear sign of the increasing
metastability is the increase in the free-energy barriers.  For the
reaction coordinates  $\xi(\theta)=q_1$, $\xi(\theta)=\beta$, 
$\xi(\theta)=-\log\{p(\theta)p(y|\theta)\}$, we had to run the adaptive
algorithm for $T=10^9$ iterations in order to obtain a converged bias and
recover a biased posterior sample which is symmetric by
labelling. Again, a more elaborate proposal strategy for the
Hastings-Metropolis step in the adaptive algorithm would be likely to
stabilize the bias faster.  As an illustration, Figure
\ref{fig:Hidalgo_beta} presents the trajectories of
$(\mu_1,\mu_2,\mu_3)$ sampled by the adaptive algorithm, with random
walk scales: $\tau_q = 0.001$, $\tau_\mu = 0.05$, $\tau_v = 0.1$ and
$\tau_\beta = 0.005$.  The trajectories should be compared to the ones
depicted in Figure~\ref{fig:no_bias}, which are obtained with the
same proposal, but without any biasing procedure.

\begin{figure}[htbp]
  \center
  \includegraphics[width=4cm,angle=270]{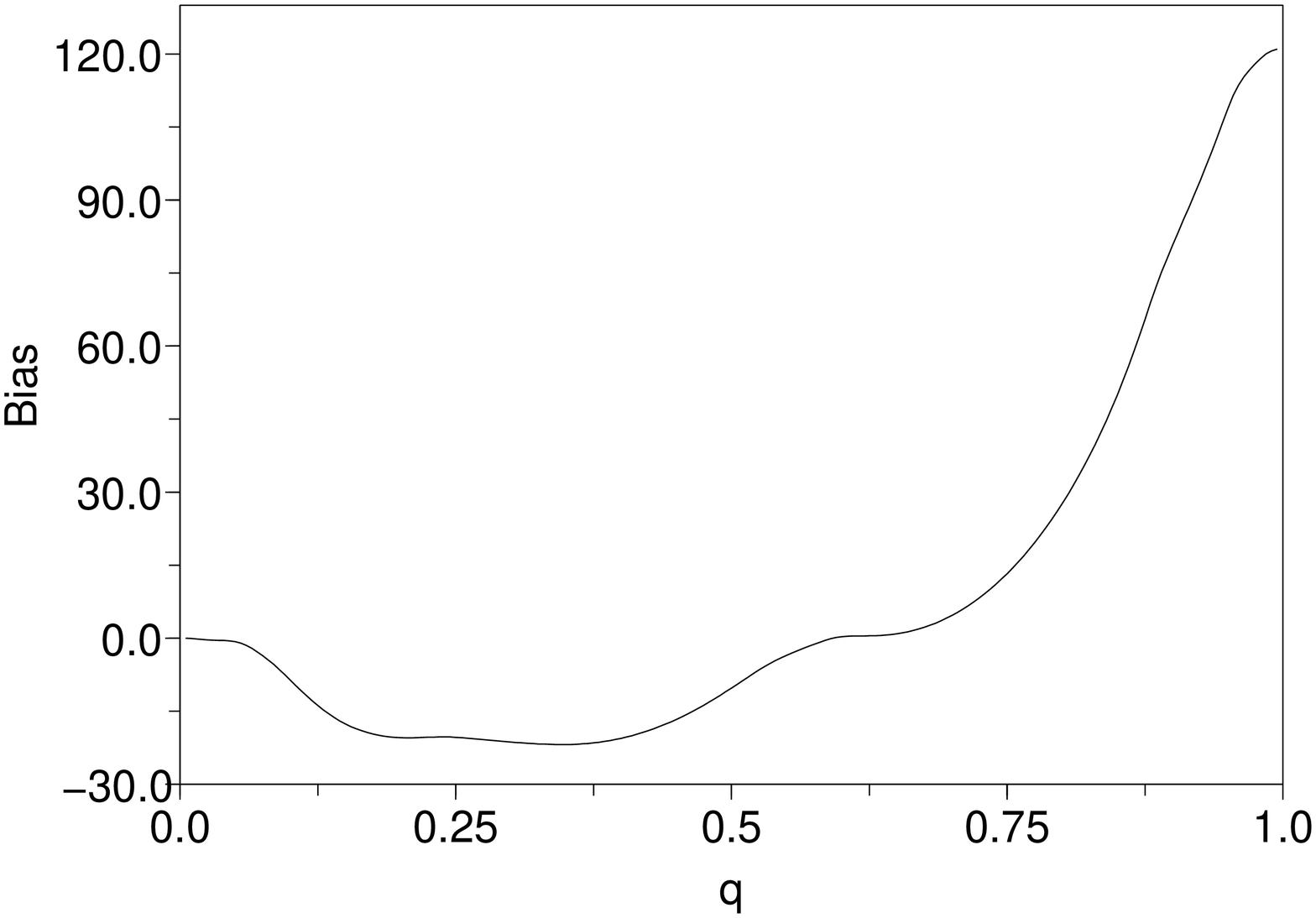}
  \includegraphics[width=4cm,angle=270]{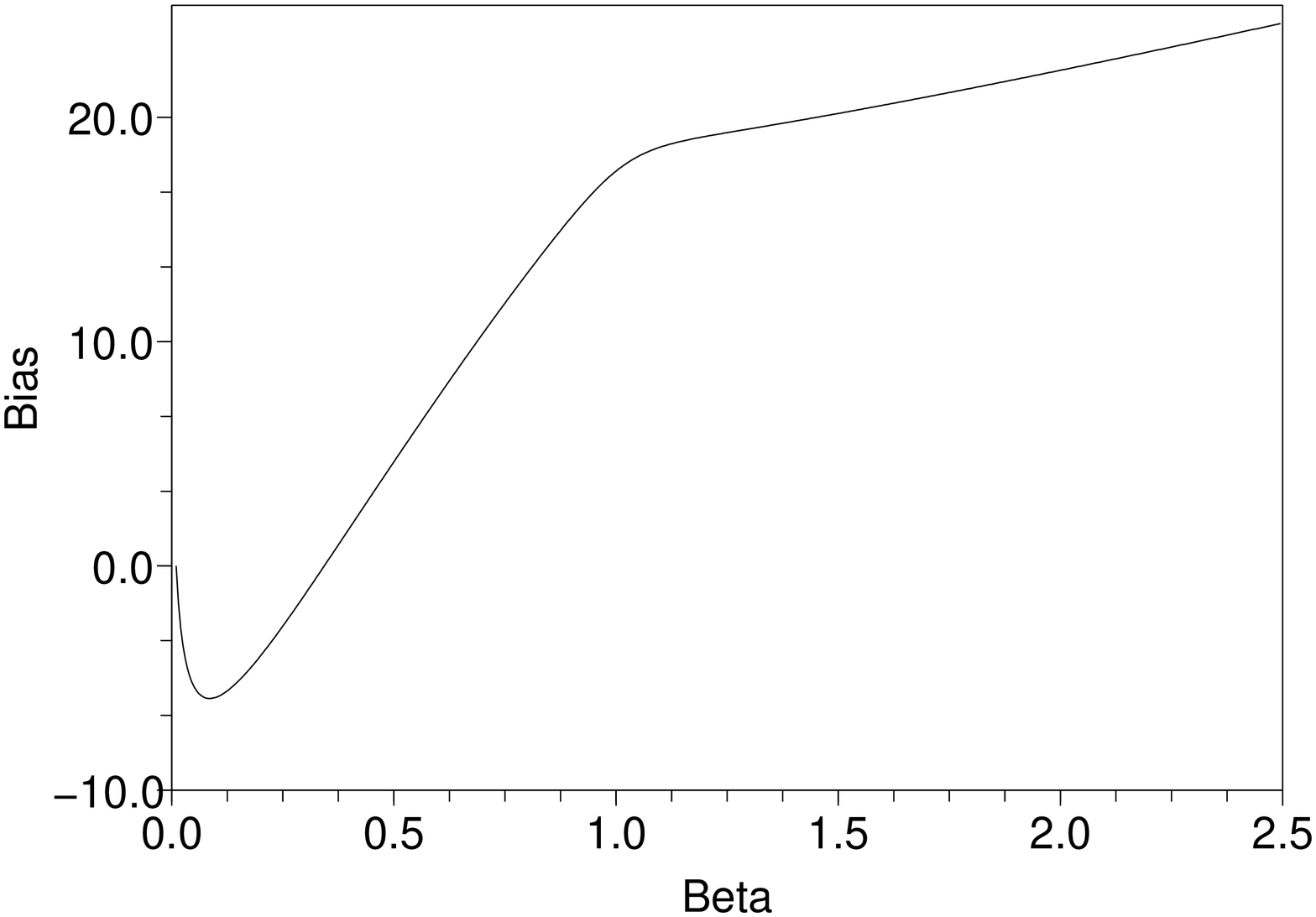} \\
  \includegraphics[width=4cm,angle=270]{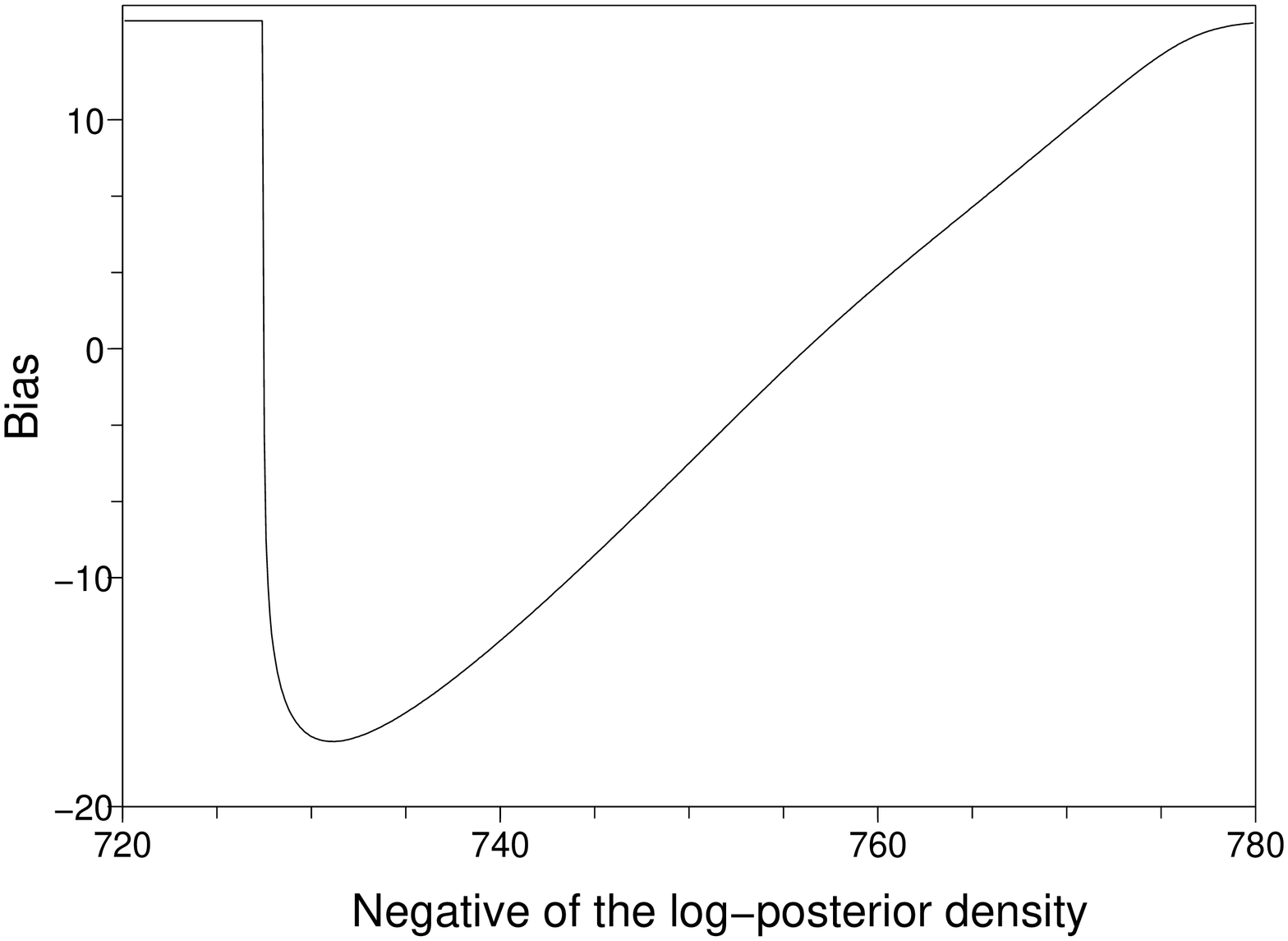}
  \caption{\label{fig:hidalgo_bias} 
    Hidalgo stamps problem. Free
    energies obtained for the reaction coordinates:
    $\xi(\theta)=q_1$ (top left), $\xi(\theta)=\beta$ (top right) and
    $\xi(\theta)=-\log\{p(\theta)p(y|\theta)\}$ (bottom).}
\end{figure}

In Figure~\ref{fig:hidalgo_bias}, we represent the biases obtained
with various choices of the reaction coordinate.  In the case
$\xi(\theta) = \beta$, we set $z_{\rm min} = 0.005$,
$z_{\rm max} = 2.5$ and $\Delta z = 0.005$. For $\xi(\theta) = q_1$,
we consider $z_{\rm min} = 0$, $z_{\rm max} = 1$ and
$\Delta z = 0.005$.  Finally, for
$\xi(\theta)=-\log\{p(\theta)p(y|\theta)\}$, we choose
$z_{\rm min} = 720$, $z_{\rm max} = 780$ and $\Delta z = 0.1$.  

\begin{figure}[htbp]
  \center
  \includegraphics[width=7.3cm]{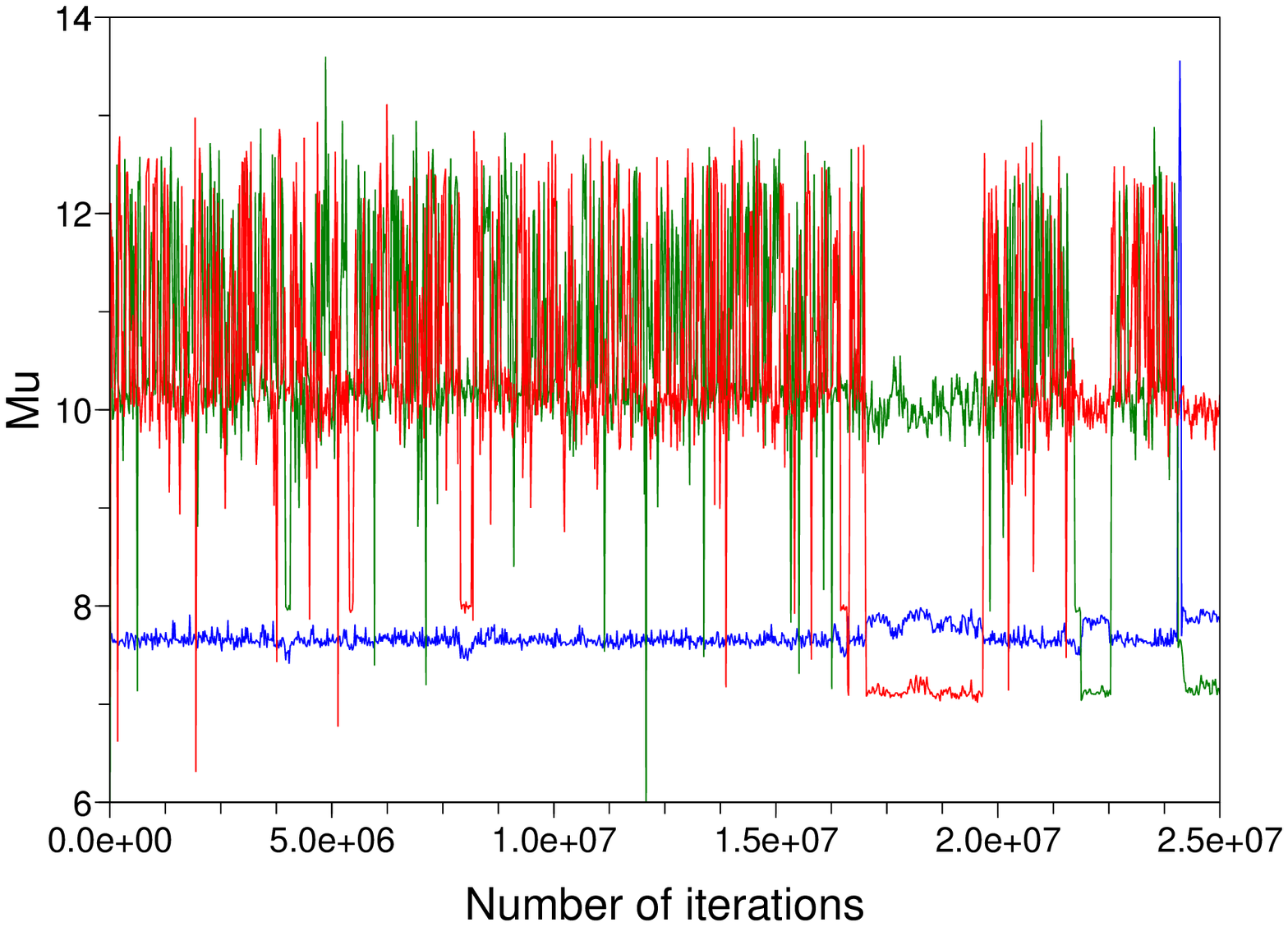}
  \includegraphics[width=7.3cm]{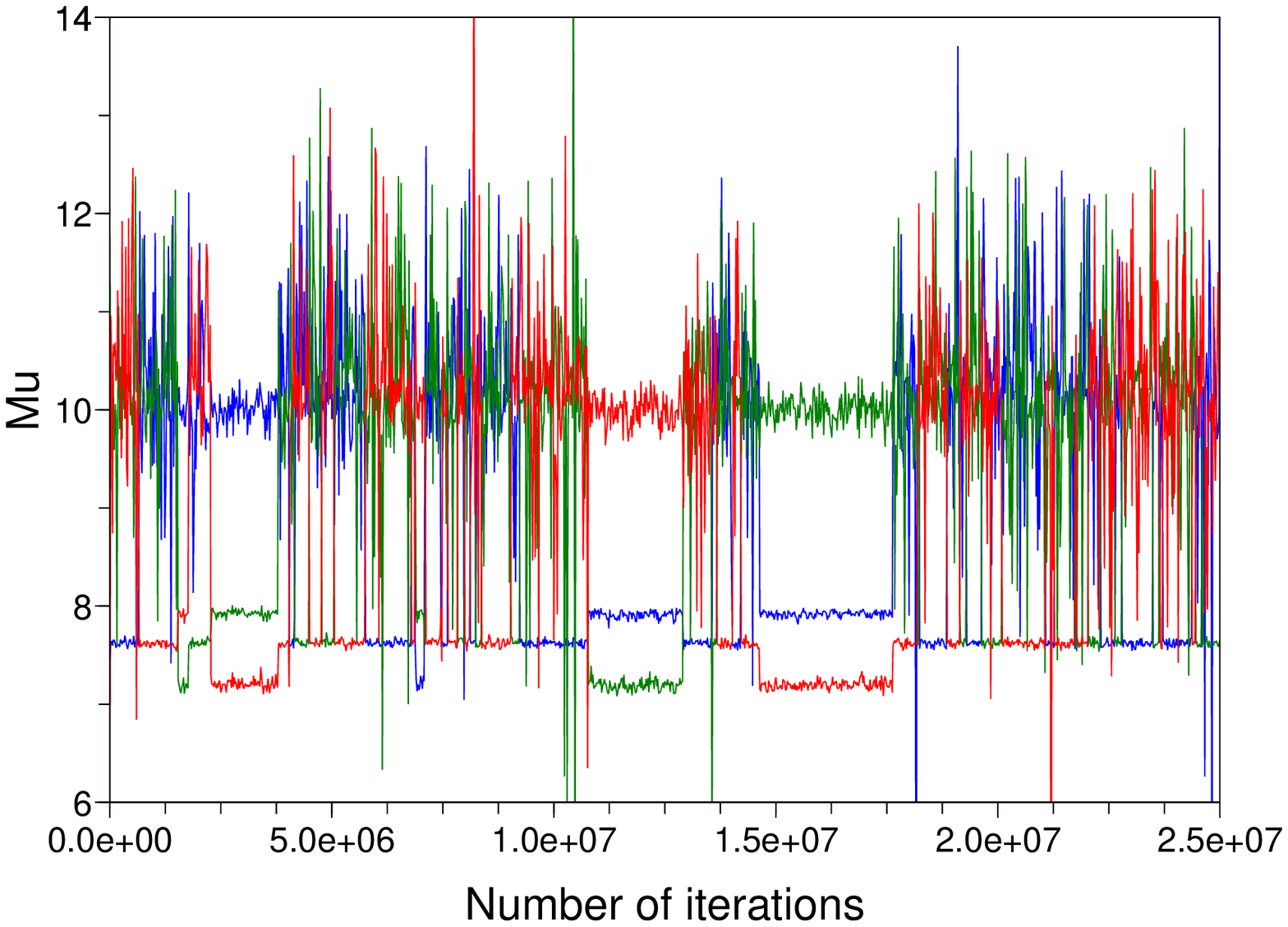}\\
  \includegraphics[width=7.3cm]{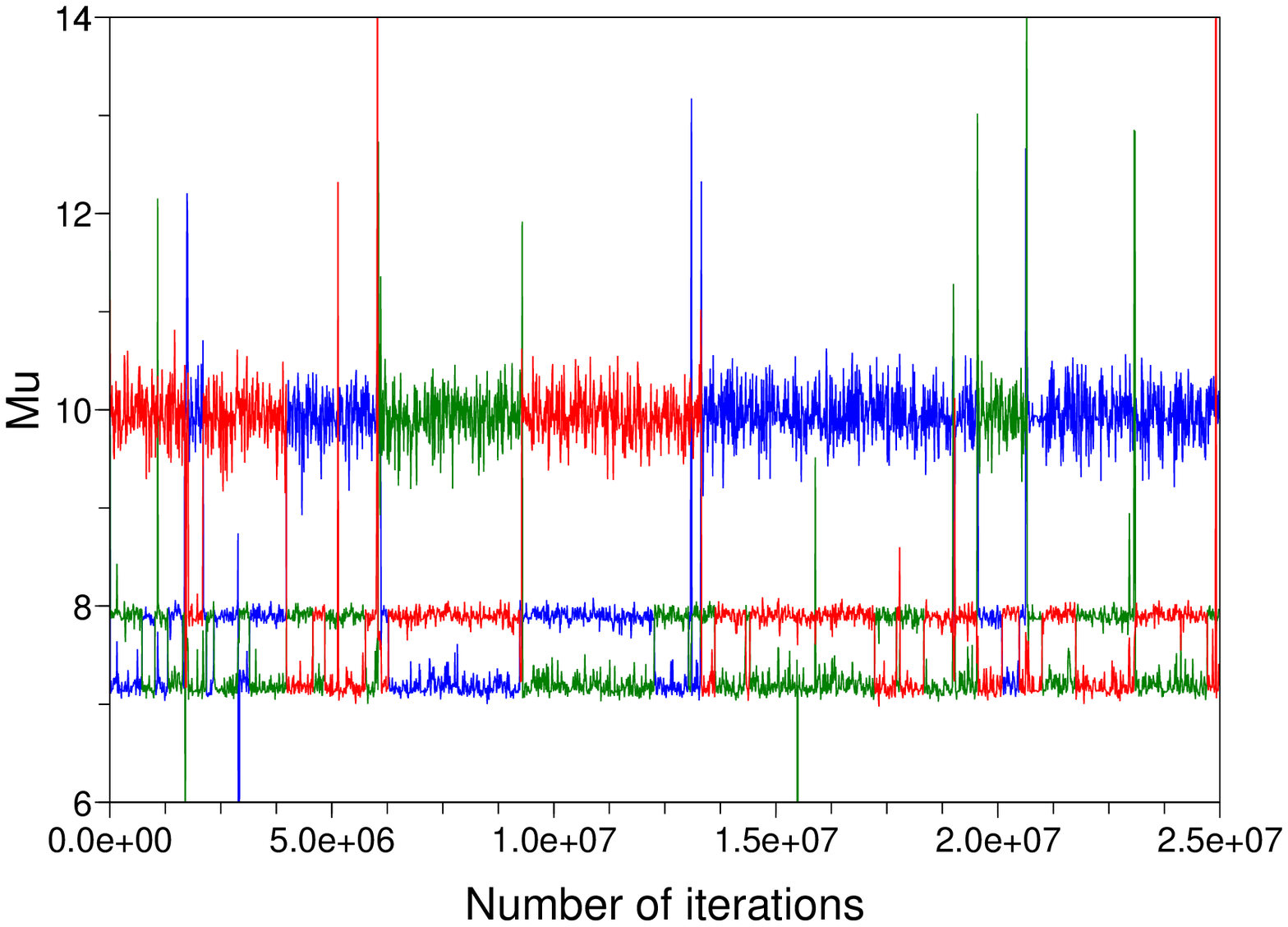}
  \caption{\label{fig:hidalgo_bias_traj} Hidalgo stamps problem. 
    Trajectories of $(\mu_1,\mu_2,\mu_3)$ of 
    the biased dynamics. Top left: reaction coordinate $\xi(\theta) = q_1$. 
    Top right: $\xi(\theta)=\beta$.
    Bottom: $\xi(\theta)=-\log\{p(\theta)p(y|\theta)\}$.
    }
\end{figure}

Biased trajectories are presented in
Figure~\ref{fig:hidalgo_bias_traj} for $\xi(\theta)=q_1$, $\xi(\theta)=\beta$, 
and $\xi(\theta)=-\log\{p(\theta)p(y|\theta)\}$. 
Efficiency factors are reported in Table~\ref{tab:hidalgo_EF_RC}.
The results show that, in terms of mode switching, 
$\xi(\theta)=-\log\{p(\theta)p(y|\theta)\}$ is the best choice.
The choice $\xi(\theta) = q_1$, although it leads to the highest efficiency factor, 
is a poor choice 
since very few switchings are observed; in particular, the mode starting around~7.5
does not change during the first $2.5 \times 10^7$ iterations.
Such transitions are observed in the case $\xi(\theta) = \beta$.

\begin{table}
\begin{center}
\begin{tabular}{c|c|c|c|c}
Reaction coordinate & $\beta$ & $-\log\{p(\theta)p(y|\theta)\}$ & $q_1$ \\
\hline
EF (numerical)   & 0.02  & 0.24  & 0.23  \\
EF (theoretical) & 0.06 & 0.13 & 0.18 \\   
\end{tabular}
\end{center}
\caption{Efficiency factor for various choices of reaction coordinates, in the case
$K=3$.}
\label{tab:hidalgo_EF_RC}
\end{table}

\section{Conclusion}
\label{sec:conclusion}

We showed in this paper how to sample efficiently posteriors of
univariate Gaussian mixture models, using a free energy biasing approach, which can be 
summarized as follows:
\begin{enumerate}[\quad (1)]
\item We choose a reaction coordinate (a function $\xi$ of $\theta$). 
\item We run an \emph{adaptive} MCMC sampler, in order
  to compute an estimation of the free energy~$A$ associated to $\xi$. 
\item From the estimated free energy $\widehat{A}$ (the
  output of the previous step), we define
  the biased density $\widetilde{\pi}=\pi_{\widehat{A}}$. We run a \emph{standard} MCMC sampler
  that targets $\widetilde\pi$ (say a Gaussian random-walk Hastings-Metropolis algorithm).
\item We do an importance sampling step, from $\widetilde{\pi}$ to
  $\pi$ to remove the bias and recover the true
  posterior~$\pi$. 
\end{enumerate}
In the particular case of univariate Gaussian
mixture models, a good choice for the reaction coordinate is~$\xi=\beta$ or 
$\xi(\theta)=-\log\{p(\theta)p(y|\theta)\}$. 
When $\beta$ is chosen, it is easy to estimate an interval of typical values for $\beta$ as
$[c_1R^2,c_2R^2]$ (with~$R$ the
range of the data, and $c_1 < c_2$ small constants, say $c_1=1/2000$ and $c_2=1/20$), 
while the determination of such an interval for $-\log\{p(\theta)p(y|\theta)\}$
does not seem to be straightforward.

\begin{figure}[htbp]
  \center
  \includegraphics[width=14cm]{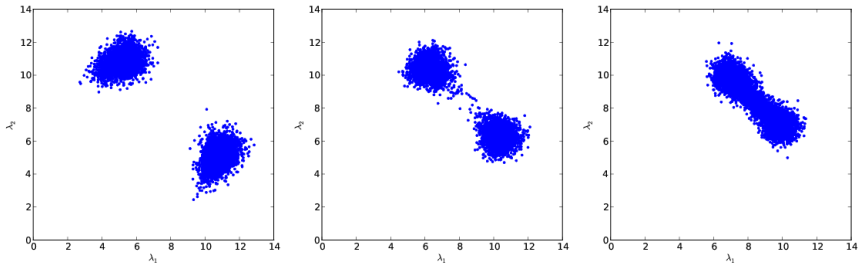}
  \caption{\label{fig:mixpoi} Scatter plots of 1000 simulated pairs
    $(\lambda_1,\lambda_2)$ from the posterior distribution of a
    two-component Poisson mixture model, and $n=100$ simulated data points,
    with a $\gam(\beta\bar{y},\beta)$ prior for the $\lambda_k$, and, from left to right, $\beta=1$, 10,
    20.}
\end{figure}

We think that the same ideas may be applied to other mixture models. For instance,
Figure \ref{fig:mixpoi} plots the posterior density of a two-component
Poisson mixture model, conditional on different values for the
hyper-parameters.  Specifically,
$p(y_i|\theta)=q_1\poi(y_i;\lambda_1)+(1-q_1)\poi(y_i;\lambda_2)$ for
$i=1,\ldots,n$, where $\poi(\cdot;\lambda)$ denotes the probability
density function of a Poisson distribution of parameter~$\lambda$. 
We use a $\gam(\beta\bar{y},\beta)$
prior for the $\lambda_k$'s, and a uniform prior for $q_1$. The
$n=100$ observations are simulated from this model with parameters
$(q_1,\lambda_1,\lambda_2)=(0.7,3,10)$. It can be seen again that biasing the posterior
distribution towards larger values of $\beta$ makes it possible to
reduce the distance between the different modes. 
We also obtained interesting preliminary results for multivariate
Gaussian mixtures.

We would like to highlight some practical advantages of our
approach. First, it requires little tuning: The main tuning parameters are the
scales of the random walks in both algorithms (adaptive, and MCMC),
and we obtained satisfactory results without trying to optimize
these scales. Second, it is easy to check that the final results are
correct: If the free energy has been well estimated, and the MCMC
algorithm for the biased posterior has converged, then
a nearly uniform marginal distribution for the reaction
coordinate is observed, and the marginal distributions for the
$(\mu_k,\lambda_k,q_k)$ are nearly identical, because of the symmetry of the true
posterior, and the numerous mode switchings in the MCMC trajectories.

Finally, one natural question is how to extend such an approach to
other classes of Bayesian models. As we have made clear already, the choice of
the reaction coordinate is the crucial point. If the reaction
coordinate is poorly chosen, a free energy biasing approach will bring
no benefit. As for applications in computational Statistical Physics, there 
is no general recipe for choosing this reaction
coordinate. Nonetheless, the following simple remarks may be considered as some guidelines. First,
it seems worth investigating alternatives to the reaction
coordinate usually chosen in Statistics, namely the negative of the posterior
log-density. Second, in doing so, one may keep in mind the
interpretation we proposed for $\xi(\theta)=\beta$ in Section
\ref{sec:react-coord-mixt}, \emph{i.e.} a particular parameter that
fixes to some extent the size of the energy barriers between the
different modes. In particular, in a given Bayesian hierarchical
model, the hyper-parameters at the highest level of the hierarchy
could be interesting candidates, because their values strongly influence the typical
values of the other components of the system.
More research in this direction is however required to draw more
definite conclusions.


\subsection*{Acknowledgements}

Part of this work was done while the two last authors were
participating to the program ``Computational Mathematics'' at the
Haussdorff Institute for Mathematics in Bonn, Germany. Support from
the ANR grants ANR-008-BLAN-0218 and ANR-09-BLAN-0216 of the French
Ministry of Research is acknowledged. The authors thank Julien
Cornebise, Arnaud Doucet, Peter J. Green, Pierre Jacob, Christian
P. Robert, Gareth Roberts, and the referees for insightful remarks.


\end{document}